\documentclass[a4paper,11pt]{article}
\pdfoutput=1

\usepackage{jcappub} 
\usepackage[T1]{fontenc} 
\usepackage{multirow}
\usepackage{array, makecell}
\usepackage{float} 
\usepackage{tabu}

\newcommand{\be}{\begin{eqnarray}}

\title{\boldmath Baryonic effects for weak lensing.\\Part II. Combination with X-ray data and extended cosmologies}

\author[a,b]{Aurel Schneider,}
\author[b]{Alexandre Refregier,}
\author[c]{Sebastian Grandis,}
\author[d]{Dominique Eckert,}
\author[b]{Nicola Stoira,}
\author[b]{Tomasz Kacprzak,}
\author[a]{Mischa Knabenhans,}
\author[a]{Joachim Stadel,}
\author[a]{and Romain Teyssier\,}

\affiliation[a]{Institute for Computational Science, University of Zurich, Winterthurerstrasse 190, 8057 Zurich, Switzerland}
\affiliation[b]{Institute for Particle Physics and Astrophysics, ETH Zurich, Wolfgang Pauli Strasse 27, 8093 Zurich, Switzerland}
\affiliation[c]{Faculty of Physics, Ludwig-Maximilians-Universit\"at, Scheinerstrasse 1, 81679 Munich, Germany}
\affiliation[d]{Department of Astronomy, University of Geneva, Ch. d'Ecogia 16, 1290 Versoix, Switzerland}

\emailAdd{aurel.schneider@uzh.ch}

\abstract{An accurate modelling of baryonic feedback effects is required to exploit the full potential of future weak-lensing surveys such as Euclid or LSST. In this second paper in a series of two, we combine Euclid-like mock data of the cosmic shear power spectrum with an eROSITA X-ray mock of the cluster gas fraction to run a combined likelihood analysis including both cosmological and baryonic parameters. Following the first paper of this series, the baryonic effects (based on the \emph{baryonic correction model} of Ref.~\citep{Schneider:2018pfw}) are included in both the tomographic power spectrum and the covariance matrix. However, this time we assume the more realistic case of a $\Lambda$CDM cosmology with massive neutrinos, and we consider several extensions of the currently favoured cosmological model. For the standard $\Lambda$CDM case, we show that including X-ray data reduces the uncertainties on the sum of the neutrino mass by $\sim30$ percent, while there is only a mild improvement on other parameters such as $\Omega_m$ and $\sigma_8$. As extensions of $\Lambda$CDM, we consider the cases of a dynamical dark energy model (wCDM), a $f(R)$ gravity model (fRCDM), and a mixed dark matter model ($\Lambda$MDM) with both a cold and a warm/hot dark matter component. We find that combining weak lensing with X-ray data only leads to a mild improvement of the constraints on the additional parameters of wCDM, while the improvement is more substantial for both fRCDM and $\Lambda$MDM. Ignoring baryonic effects in the analysis pipeline leads to significant false-detections of either phantom dark energy or a light subdominant dark matter component. Overall we conclude that for all cosmologies considered, a general parametrisation of baryonic effects is both necessary and sufficient to obtain tight constraints on cosmological parameters.}

\begin{document}
\maketitle

\section{Introduction}
The main goal of cosmological weak-lensing surveys is to learn more about our standard model of cosmology $\Lambda$CDM (and potential extensions thereof) by measuring the total matter distribution of the universe. In recent years, however, it has become clear that astrophysical feedback processes significantly affect the cosmological density field, making it more difficult to extract information about fundamental physics. Very energetic feedback processes attributed to active galactic nuclei (AGN) are believed to push large amounts of gas into the intergalactic space, which causes up to 10-30 percent changes in the clustering signal at small and medium cosmological scales \citep{vanDaalen:2011xb,Chisari:2019tus}. The exact amplitude of this effect, however, remains unknown, since AGN feedback is caused by processes around super-massive black holes that cannot be simulated self-consistently at cosmological scales.

Current cosmological hydrodynamical simulations include AGN feedback effects as semi-analytical sub-grid models that are tuned to observations. Since different simulations rely on different sub-grid prescriptions and are calibrated to different observations, they do not agree in terms of the cosmological weak-lensing signal. While simulations such as {\tt CosmoOWLS} \citep{Mummery:2017lcn}, {\tt Illustris} \citep{Vogelsberger:2014dza}, or {\tt BAHAMAS} \citep{vanDaalen:2019aaa} predict a 20-30 percent baryonic suppression effect of the clustering signal, other simulations like {\tt EAGLE} \citep{Hellwing:2016ucy}, {\tt HorizonAGN} \citep{Chisari:2018prw}, or {\tt Illustris-TNG} \citep{Springel:2017tpz} find a weaker effect of order 10 percent or less.

Given that the amplitude of the baryonic effects on the clustering signal is uncertain, one way forward is to parametrise the problem, thereby adding further nuisance parameters to the cosmological analysis pipeline. Such an approach can be pursued using fitting functions \citep{Harnois-Deraps:2014sva}, the halo model \citep{Semboloni:2011aaa,Fedeli:2014gja,Mohammed:2014mba,Mead:2015yca,Debackere:2019cec}, or full hydrodynamical simulations \citep{Brun:2013yva,McCarthy:2017csu}, depending on the analysis statistic and the precision requirements. Another option is to modify gravity-only $N$-body simulations to mimic baryonic effects on the cosmological density field. Such a \emph{baryonification} approach, that is build upon a physically motivated parametrisation, has recently been adopted in Refs.~\citep{Schneider:2015wta,Schneider:2018pfw}.

The parametrisation of the \emph{baryonification} method (i.e. the baryonic correction model) is based on halo profiles and therefore related to the halo model approach. However, in contrast to the halo model, the parametrised halo profiles are used to define a particle displacement field which is applied to outputs of $N$-body simulations. This means that the baryonic correction model is not restricted to the power spectrum as output statistics but provides a modified realisation of the total density field based on the original $N$-body simulation. Recent examples of alternative applications are weak-lensing peak statistics from convergence maps \citep{Weiss:2019jfx} or a neural network approach for cosmological inference \citep{Fluri:2019qtp}.

Compared to full hydrodynamical simulations, the baryonic correction model of \citet[henceforth S19]{Schneider:2018pfw} has the advantage of providing a clearly defined parametrisation of baryonic effects and the possibility to explore this parameter space without running expensive new simulations. However, it consists of an approximative method that has to be validated. In S19 we showed that the model is able to reproduce the matter power spectrum of a variety of different simulations at the level of 2 percent or better if the baryonic model parameters are fitted to the cluster gas fractions of the same simulation. This level of accuracy is very encouraging, allowing us to use the method for cosmological parameter estimates.

The present paper consists of the second paper in a series of two dedicated to a forecast analysis for stage-IV weak-lensing surveys such as Euclid\footnote{\texttt{https://www.euclid-ec.org/}} or LSST\footnote{\texttt{https://www.lsst.org/}}. In the first paper \citep[Paper I]{Schneider:2019snl} we set up a mock observable consisting of the tomographic shear power spectrum and covariance matrix. We developed a \emph{baryonic emulator} based on the \emph{baryonic correction model} of S19 allowing us to speed up the prediction pipeline, making it fit for Markov Chain Monte Carlo sampling of the high-dimensional cosmological and baryonic parameter space. With this at hand, we carried out a first forecast analysis assuming a $\Lambda$CDM model with five cosmological parameters ($\Omega_m$, $\Omega_b$, $\sigma_8$, $h_0$, and $n_s$) where neutrino masses ($\Sigma m_{\nu}$) were set to zero for simplicity.

The main conclusions of Paper I can be summarised as follows: (i) ignoring baryonic effects in the prediction pipeline leads to strong biases of 5-10 standard deviations on all cosmological parameters except $\Omega_b$ which cannot be well constrained by weak-lensing; (ii) ignoring the small cosmological scales (modes above $\ell=100$) makes the biases disappear but leads to a strong increase of expected errors on cosmological parameters; (iii) fixing baryonic parameters instead of marginalising over them reduces the parameter contours by $\sim50$ percent or less for $\Omega_m$, $\sigma_8$ and $h_0$, while it is more than a factor of two for $n_s$; (iv) no less and no more than three baryonic parameters are required to obtain converged posterior contours for cosmology. Next to these main findings of Paper I, we furthermore showed that it is save to ignore both baryonic effects on the covariance matrix as well as the cosmology-dependence of baryonic effects on the power spectrum.

In this second paper of the series (Paper II), we will extend our analysis to a $\Lambda$CDM model including massive neutrinos. Furthermore, we will also include a mock data-set of cluster gas fractions from the X-ray survey eROSITA. This allows us to quantify how much can be gained in terms of cosmology if the baryonic parameters are further constrained using an external data-set of direct gas observations. Finally, we will go beyond the standard model of cosmology and study three extensions to $\Lambda$CDM. These include the dynamical dark energy model wCDM, the $f(R)$ modified gravity model fRCDM, and a mixed dark matter model $\Lambda$MDM, where the dark matter sector is assumed to consist of a cold and a warm/hot component.

The paper is organised as follows: Sec.~\ref{sec:predictions} provides a summary of the prediction pipeline including the most important aspects of the baryonic correction model and the calculation of the tomographic shear power spectrum. Sec.~\ref{sec:mocks} is dedicated to the mock observables. We thereby review the setup of the weak-lensing mock and explain how we construct mock observations of the mean cluster gas fraction based on the eROSITA X-ray survey. In Sec.~\ref{sec:likelihoodanalysis} we present the results from our parameter inference analysis for $\Lambda$CDM (with massive neutrinos), while extensions of the $\Lambda$CDM model are discussed in Sec.~\ref{sec:extensions}. Sec.~\ref{sec:conclusions} provides a final summary of the results, and in Appendix~\ref{app:fixedbaryons} we present results for the idealistic case when baryonic parameters are fixed to their true values used in the mock data.

\section{Summarising the prediction pipeline}\label{sec:predictions}
A fast and sufficiently accurate prediction pipeline is essential to perform Markov Chain Monte Carlo (MCMC) sampling of the cosmological parameter space. In this section we provide a broad summary of our predictions for the tomographic shear power spectra. This includes the gravity-only matter power spectrum, the baryonic effects, corrections for the presence of massive neutrinos, the selected redshift binning, the Limber approximation etc. For a more detailed description, we refer to Paper I \citep{Schneider:2019snl}.

\subsection{Baryonic correction model}\label{sec:BCM}
The baryonic correction model developed in Refs.~\citep{Schneider:2015wta,Schneider:2018pfw} consist of the backbone of this study. The task of the model is to \emph{baryonify} outputs of gravity-only $N$-body simulations by slightly displacing simulation particles in a post-processing way. The particle displacements transform the original NFW profiles of haloes ($\rho_{\rm nfw}$) into a baryon-corrected profile ($\rho_{\rm bcm}$) that accounts for the effects of stars and gas:
\begin{equation}\label{profiles}
\rho_{\rm nfw}(r) \hspace{0.5cm}\longrightarrow\hspace{0.5cm}\rho_{\rm bcm}(r) =  \rho_{\rm clm}(r) + \rho_{\rm gas}(r) + \rho_{\rm cga}(r).
\end{equation}
The profiles $\rho_{\rm gas}$, $\rho_{\rm cga}$, and $\rho_{\rm clm}$ correspond to the gas, the central galaxy, and the collisionless matter components (consisting of dark matter, satellite galaxies, and intra-cluster stars), which are parametrised so that they agree with observations. Here we only discuss the gas profile, which is given by
\begin{equation}\label{rhogas}
\rho_{\rm gas} \propto \frac{1}{(1+r/r_{\rm co})^{\beta}[1+(r/r_{\rm ej})^2]^{(7-\beta)/2}},\hspace{1cm}\beta=3-\left(\frac{M_c}{M}\right)^{\mu},
\end{equation}
where $r_{\rm co}=0.1\times r_{200}$ and
\begin{equation}\label{thetaej}
r_{\rm ej}=\theta_{\rm ej}\times r_{200}.
\end{equation}
The gas profile is characterised by a central core (of size $r_{\rm co}$) followed by a power law-decline (with slope $\beta$) and a steep truncation at the maximum ejection radius ($r_{\rm ej}$). Note that the slope of the profile described by the $\beta$ parameter depends on the halo mass, becoming shallower for smaller haloes. This mass-dependence is in agreement with observations from X-ray profiles and can be qualitatively explained by the fact that the efficiency of feedback processes depends on the underlying gravitational potential.

The gas profile has three free parameters, one related to the maximum gas ejection ($\theta_{\rm ej}$) and two related to the slope ($M_c$, $\mu$). These are the baryonic parameters that we allow to vary together with the cosmological parameters. The remaining two parameters of the \emph{baryonic correction} model ($\eta_{\rm star}$, $\eta_{\rm cga}$) describe the stellar abundances in the halo and are fixed for simplicity. In Paper I we have shown that including $\eta_{\rm star}$ and $\eta_{\rm cga}$ as free model parameters has no noticeable effect on the cosmological parameter contours.

Regarding the matter power spectrum, the baryonic effects can be quantified via the ratio $S_{\rm BCM}\equiv P_{\rm dmb}/P_{\rm dmo}$, where the dark-matter-only ($P_{\rm dmo}$) and dark-matter-baryon ($P_{\rm dmb}$) matter power spectra are measured directly from the particle data. While the dark-matter-only data corresponds to the $N$-body simulation output, the data including baryonic effects is previously modified using the \emph{baryonionification} method described above. All measurements of the power spectrum are performed with the internal power spectrum calculator of {\tt Pkdgrav3} \citep{Stadel:2001aaa,Potter:2016ttn}.

In Paper I we have developed a \emph{baryonic emulator} that allows to obtain fast predictions of the ratio $S_{\rm BCM}$ suitable for MCMC sampling. The emulator was developed with the uncertainty quantification software {\tt UQLab} \citep{Marelli:2014aaa} that relies on polynomial chaos expansion as spectral decomposition method. The code generates a surrogate model which can be evaluated at any point in the multidimensional parameter space. The baryonic emulator includes five baryonic parameters ($M_c$, $\mu$, $\theta_{\rm ej}$, $\eta_{\rm star}$, $\eta_{\rm cga}$) and one cosmological parameter ($f_b=\Omega_b/\Omega_m$) consisting of the cosmic baryon fraction. However, as mentioned above, we fix the stellar parameters to their benchmark values $\eta_{\rm star}=0.32$ and $\eta_{\rm cga}=0.6$ (see Paper I and S19).

The baryonic correction model has been shown to be in good agreement with hydrodynamical simulations. In S19 we fitted the BC model parameters to the measured gas fractions of galaxy groups and clusters of seven different hydrodynamical simulations. We then predicted the power spectrum with the BC model, and compared the result to the same simulation. In this way we obtained an agreement of 2 percent or better up to modes of $k=10$ h/Mpc at redshift zero, validating the \emph{baryonification} approach.

The \emph{baryonic emulator}, on the other hand, has an uncertainty of 3 percent at the 2-$\sigma$ level. However, for most $k$-modes the agreement is significantly better than this. A summary plot showing the accuracy of baryonic emulator can be found in Fig.~4 of Paper I.

\subsection{Angular power spectra}
The tomographic auto and cross power spectra of the cosmic shear are obtained using the Limber approximation \citep{Limber:1953aaa} summarised in Eqs.~(3.1) and (3.2) of Paper I. For the galaxy distribution we also follow Paper I and assume
\begin{equation}\label{nofz}
n(z)\propto z^{2}\exp(-z/0.24),\hspace{1cm}z_{\rm bin}=\left[0.1,0.478,0.785,1.5\right],
\end{equation}
where $z_{\rm bin}$ defines the edges of the three tomographic bins. The calculation of the matter power spectrum, on the other hand, differs from the one used in Paper I. First, we now rely on the Boltzmann solver {\tt Class} \citep{Lesgourgues:2011re,Blas:2011rf} instead of the fit from \citet{Eisenstein:1998aaa} for the initial transfer function. While being more accurate, this allows us to include cosmologies with massive neutrinos and dynamical dark energy into the prediction pipeline. Second, we now model the nonlinear effects of massive neutrinos following the {\tt halofit} extension of \citet{Bird:2012aaa}. Note, however, that contrary to the original work of Ref~\citep{Bird:2012aaa} where the neutrino corrections have been applied to the original {\tt halofit} model of \citet{Smith:2002dz}, we apply the corrections to the {\tt revised halofit} version from \citet{Takahashi:2012em}. The relative effects of massive neutrinos on the matter power spectrum are shown in the left-hand panel of Fig.~\ref{fig:contourneutrino}.

The present analysis is limited to the normal hierarchy with one massive and two massless neutrino flavour states. While this consists of a simplifying choice, we want to stress that the expected sensitivity of a stage-IV weak lensing survey alone is not high enough to differentiate between normal and inverted neutrino mass hierarchies.

Following the model of Ref. \citep[][]{Hirata:2004aaa}, we also include a correction for the intrinsic alignment of galaxy shapes. We thereby restrict ourselves to one additional free parameter ($A_{\rm IA}$) describing amplitude of the intrinsic-intrinsic and intrinsic-shear contributions to the power spectrum. The calculation of the cosmic shear power spectrum including the intrinsic-alignment terms and the nonlinear neutrino corrections is done within the package {\tt PyCosmo} described in \citet{Refregier:2017seh}.

\begin{figure}[tbp]
\centering
\includegraphics[width=0.99\textwidth,trim=0.7cm 0.0cm 1.6cm 0.8cm,clip]{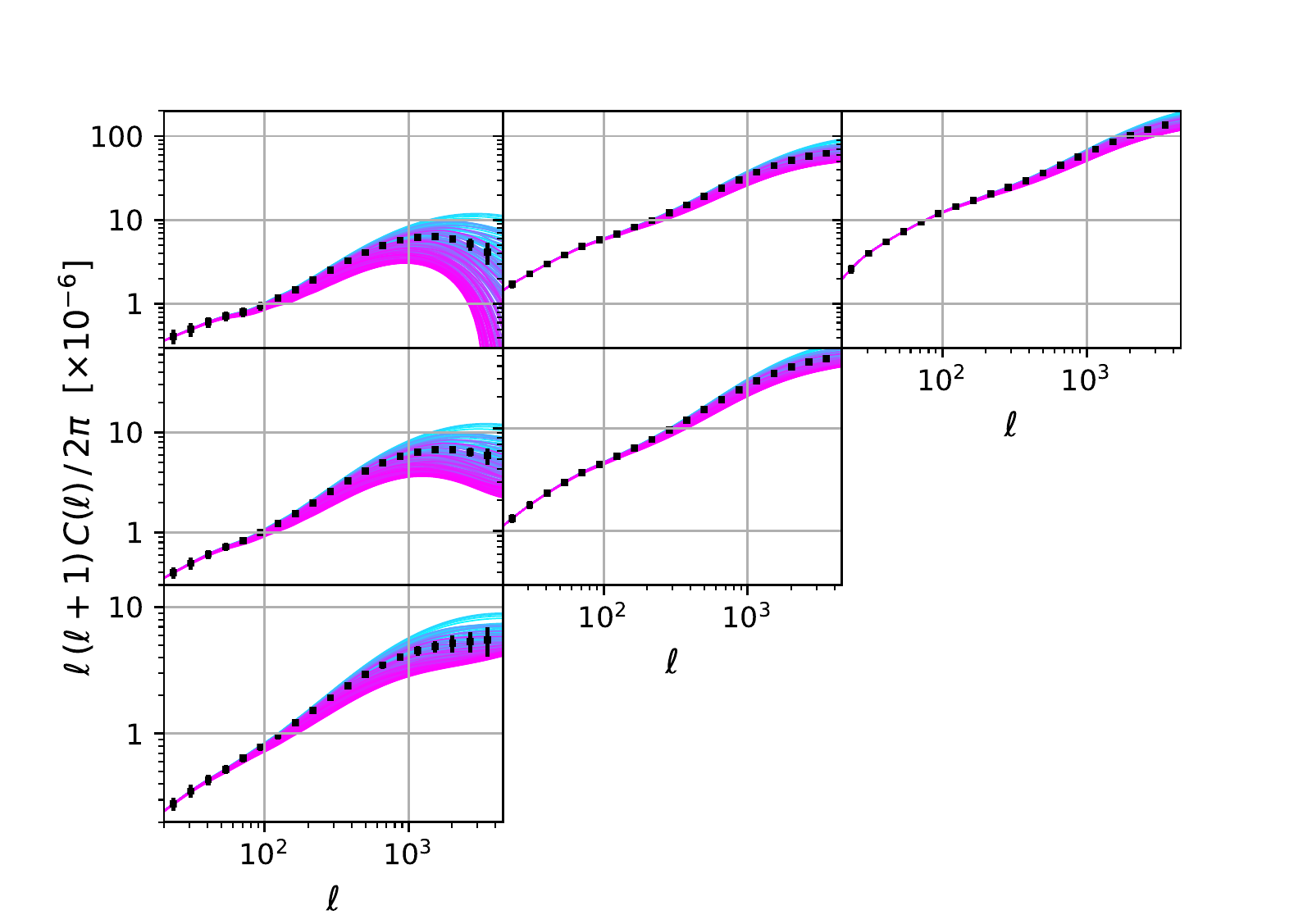}
\caption{Mock realisation of the auto and cross power spectra from three redshift bins with errors that include both sample variance and Gaussian shape noise of a Euclid-like survey (black data points). The auto power spectra are shown by the diagonal, the cross spectra by the off-diagonal panels (with decreasing redshift from top-right to bottom-left). The data is constructed from the galaxy distribution and redshift bins of Eq.~(\ref{nofz}). The coloured lines correspond to the theoretical predictions assuming a range of baryonic parameters $\log M_c=13-16$ (mass in M$_{\odot}$/h), $\mu=0.1-0.7$ and $\theta_{\rm ej}=2-8$. The spread of the lines illustrates the current level of uncertainty due the baryonic effects.}
\label{fig:Cofl}
\end{figure}

Fig.~\ref{fig:Cofl} shows the resulting cosmic shear auto and cross power spectra of the three tomographic bins defined in Eq.~(\ref{nofz}). The auto spectra are shown by the diagonal panels, the cross-spectra by the off-diagonal panels (with decreasing redshift from top-right to bottom-left). The coloured lines correspond to predictions assuming a fixed {\tt Planck} cosmology plus varying values for the three baryonic parameters ($\log M_c$, $\mu$, and $\theta_{\rm ej}$) within the prior ranges given in Table~\ref{tab:prior0}. The band of coloured lines therefore provides a measure for the current uncertainties related to baryonic effects. The black data points with error bars correspond to the mock observable which will be presented in the following section. Note that Fig.~\ref{fig:Cofl} differs from Fig.~5 of Paper I in the sense that the former is based on the Boltzmann solver {\tt Class} assuming a six parameter $\Lambda$CDM model with a neutrino of mass $m_{\nu}=0.1$ eV, while the latter is based on a 5 parameter model with massless neutrinos obtained via the less accurate \citet{Eisenstein:1998aaa} fitting function.

\section{Mock observations}\label{sec:mocks}
In this section, we present the mock measurements of the weak-lensing shear power spectrum based on a Euclid-like setup as well as the X-ray cluster gas fraction assuming future observations from eROSITA. Since the weak-lensing mock is very similar to the one from Paper I \citep{Schneider:2019snl}, we only provide a broad summary of its main characteristics. The construction of the eROSITA-based X-ray mock, on the other hand, is explained in more detail.

\subsection{Weak-lensing mock data}\label{sec:WLmock}
Our weak-lensing mock sample requires both a theoretical prediction for the tomographic auto and cross angular spectra plus a corresponding covariance matrix. The power spectra are obtained from the pipeline described in the previous section. We thereby assume the galaxy distribution and redshift binning of Eq.~(\ref{nofz}). The considered angular scales cover the range $20\leq\ell\leq4000$ with 19 logarithmically spaced data points  for each of the six auto and cross power spectra.

The covariance matrix is built from a sample of 300 Euclid-like weak-lensing maps for each of the three redshift bins (summing up to a total of 900 maps). The maps originate from 10 independent simulations with different random seed initial conditions. For each of these simulations, we construct 10 different light-cones by adding arbitrary shifts and rotations during the box replication process. All light-cones are then transformed into weak-lensing convergence maps using the Born approximation. We add three different configurations of Gaussian shape noise (based on Eq.~3.5 of Paper I) to each map, ending up with 300 (semi-) independent realisations for the full covariance matrix including all auto and cross correlations between redshift bins. Note that this is about a factor of three larger than the size of the data vector (which consists of $6\times19=114$ data points in total).

In Paper I we have checked that the number of realisations used to build our covariance matrix is sufficient to obtain reliable 68 and 95 percent posterior contours for the cosmological and baryonic parameters. Note that, in principle, a more accurate calculation of the full posterior probability can be obtained using the method discussed in \citet{Sellentin:2015waz} (which is based on a multivariate t-distribution instead of a Gaussian likelihood used here). However, we have verified that this does not affect the contours shown in the present work (for a more detailed discussion, see Sec. 4.4 of Paper I).

The default cosmological and baryonic parameter values of the mock are given in Table~\ref{tab:prior0}. As discussed above, we now assume a neutrino of mass $m_{\nu}=0.1$ eV, which is above the lower limit of 0.06 eV from solar and atmospheric neutrino experiments \citep[e.g. Ref.][]{Ahmad:2001aaa, Ahn:2006zza} but well below the current strongest limits on the neutrino mass from cosmology \citep{Aghanim:2018eyx}. The default values of the baryonic parameters are the same than in Paper I. They are selected based on the best-fitting values of current gas fractions from X-ray cluster observations \citep{Eckert:2012aaa,Eckert:2015rlr}, see benchmark model B of S19.

Note that we include a neutrino mass of $m_{\nu}=0.1$ eV in the tomographic power spectrum but not in the simulations for the covariance matrix. While this is strictly speaking inconsistent, it is unlikely to have any effect on the posterior contours. First of all, it has been shown before that fixing cosmological parameters (instead of varying them) in the covariance matrix does not noticeably affect the outcome of a likelihood analysis \citep[see e.g. Ref.][]{Kodwani:2018uaf}. In our case, we set the neutrino mass to zero which is a particular fixed value within our prior range. Second, we explicitly showed in Paper I that ignoring baryonic effects in the covariance matrix does not lead to a visible effect on the posterior contours. Since both baryonic and massive neutrino suppression effects are of similar maximum amplitude, we therefore do not expect missing neutrino masses in the covariance matrix to affect the posterior contours either. Finally, we want to stress that massive neutrinos lead to a suppression of the power spectrum, which means that ignoring them in the covariance matrix will at most lead us to overestimate (but not underestimate) the size of the posterior contours.

The mock cosmic shear power spectra are shown as black symbols in Fig.~\ref{fig:Cofl}. The error bars correspond to (the square root values of) the diagonal terms of the covariance matrix. An illustration of the normalised covariance matrix is provided in Fig.~6 of Paper I, where the effects of shape-noise and baryons are visualised. Note that, while the effects of baryons are hardly visible in the plot, the shape-noise contribution is very important in the sense that it washes out most of the mode-coupling effects from nonlinear structure formation.

\subsection{X-ray mock data}\label{sec:Xraymock}
In Paper I we have focused on a setup where posterior contours of cosmological parameters were obtained by either fixing or marginalising over the three baryonic parameters. One of the main goals of Paper II is to include additional information on the gas content of haloes instead, in order to further constrain the baryonic parameters in a realistic context.

From previous work \citep[e.g. Refs.][]{Schneider:2015wta,Mummery:2017lcn} we know that the mean X-ray gas fractions of galaxy-groups and clusters is a particularly sensitive discriminant of the baryonic suppression effect on the weak-lensing signal. We therefore construct a mock data-set for the mean gas fractions, assuming the specifics of the upcoming X-ray eROSITA telescope \citep{Merloni:2012aaa}. The eROSITA instrument on board of the Russian `Spectrum-Roentgen-Gamma' satellite is expected to observe many thousands of individual galaxy groups and clusters over a redshift range between $z=0.1$ and $z=1.5$. Together with mass estimates from weak-lensing measurements of Euclid, it will be possible to obtain observations of the gas fraction for thousands of individual objects \citep{Grandis:2018mle}. Although each individual gas fraction can only be determined with rather modest signal-to-noise, the mean gas fraction per halo mass will be known to high precision. For the sake of constraining baryonic effects on cosmology, measuring the mean gas fraction is sufficient, since the scatter has been shown to not affect the cosmological signal \citep[see Ref.][]{Schneider:2015wta}.

We set up mock measurements of the gas fractions at two radii $r_{2500}$ and $r_{500}$ to simultaneously constrain all three baryonic parameters ($\theta_{\rm ej}$, $\mu$, and $M_{c}$). This allows to break the degeneracy between $\theta_{\rm ej}$ and $M_{c}$, which is present if the model is fitted to the gas fraction at one radius alone (see Fig.~7 of S19, for example).

\begin{figure}[tbp]
\centering
\includegraphics[width=0.98\textwidth,trim=0.6cm 0.0cm 1.5cm 0.4cm,clip]{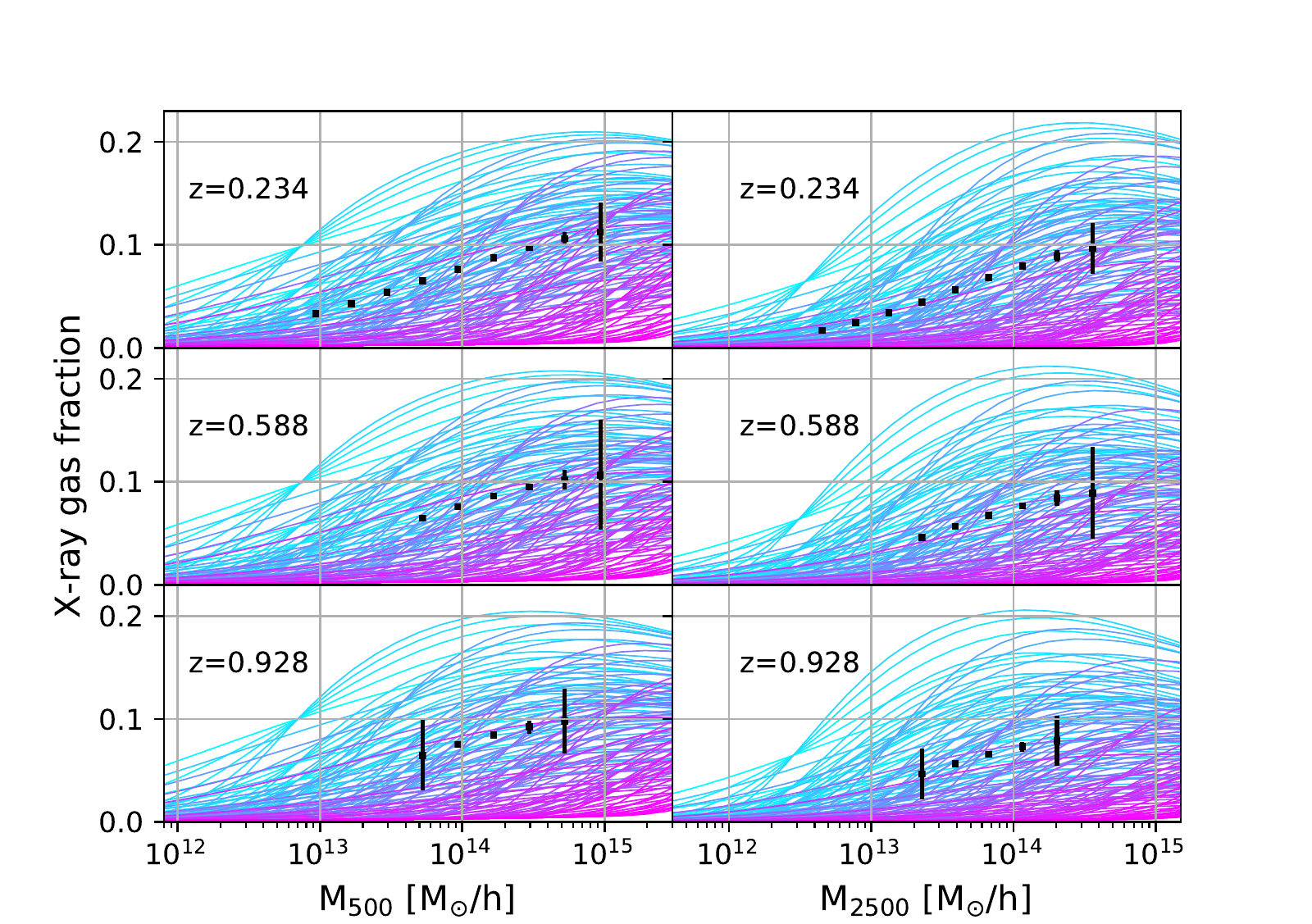}
\caption{Mock observations of the mean X-ray gas fractions of galaxy groups and clusters from eROSITA (black data points). Different rows refer to different redshifts while the left-hand and right-hand columns show the fractions at $M_{500}$ and $M_{2500}$, respectively. The coloured lines are correspond to the predictions from the baryonic correction model. They are based on the same parameter values than the lines in Fig.~\ref{fig:Cofl}.}
\label{fig:xrayfraction}
\end{figure}

The mock gas fractions are determined using the baryonic correction model with the same default parameters than what was used for the cosmic shear mock. We split the data in 9 mass-bins (designated by the subscript $i$) covering the range $M_{500}\sim10^{13}-10^{15}$  M$_{\odot}$/h and $M_{2500}\sim5\times10^{12}-5\times10^{12}$ M$_{\odot}$/h, and we assume 3 redshift-bins (designated by the subscript $k$) corresponding to the tomographic bins from the cosmic shear analysis (see Eq.~\ref{nofz}). The mass and redshift bins are populated with objects following the predicted eROSITA cluster number counts $N_{\rm cl}(M_i,z_k)$. We thereby only include the part of the sky where there is an overlap with the Euclid field-of-view \citep[see Ref.][]{Grandis:2018mle}. This is necessary because the total mass estimates are assumed to rely on cluster lensing measurements from Euclid.

The gas fractions at $M_{\Delta}$ with $\Delta$ = 500 and 2500 are given by
\begin{equation}
f_{\rm gas}(M_i,z_k)\equiv\frac{M_{\rm gas,i}(r_{\Delta},z_k)}{M_{\rm tot,i}(r_{\Delta},z_k)},
\end{equation}
where $M_{\rm gas,i}$ and $M_{\rm tot,i}$ are determined using the baryonic correction model at the mean redshifts $z_k$ = 0.238, 0.588, 0.94. The error bars on the mock gas fractions are given by
\begin{equation}\label{xrayerror}
\sigma_{f_{\rm gas}}(M_i,z_k) =\sqrt{\frac{2}{N_{cl}(M_i,z_k)\times{\rm SNR}(M_i,z_k)}+\sigma_{\rm sys}^2}
\end{equation}
where $\sigma_{\rm sys}=0.013$ is the expected systematic error on the total halo mass \citep{Grandis:2018mle}. The signal-to-noise ratio ${\rm SNR}(M_i,z_k)$ for individual haloes and the number of clusters $N_{\rm cl}(M_i,z_k)$ for each redshift bin in the context of a Euclid-type survey setup are obtained from Ref.~\citep{Grandis:2018mle} as well. Note that there is an additional factor of two in the first term on the right-hand-side of Eq.~(\ref{xrayerror}). This correction has been introduced because $f_{\rm gas}(M_{500})$ and $f_{\rm gas}(M_{2500})$ are highly correlated. An easy and conservative way to neglect the covariance between the two data vectors is to assume that we use half of the sky to measure $f_{\rm gas}(M_{500})$ and the other half to measure $f_{\rm gas}(M_{2500})$, reducing the cluster count by a factor of two.

The resulting mock data-set of the gas fractions at $r_{500}$ (left) and $r_{2500}$ (right) is shown in Fig.~\ref{fig:xrayfraction} as black data points. Due to the large numbers of expected galaxy-group and cluster detections by eROSITA, the error-bars generally remain very small. Only exceptions are the largest mass and highest redshift bins where eROSITA selected clusters and groups become rare, resulting in larger errors.

Next to the mock data set, we also show predictions of the mean gas fractions from the baryonic correction model with varying baryonic parameters $\theta_{\rm ej}$, $\mu$, and $M_{c}$ within their prior values given in Table~\ref{tab:prior0}. The resulting coloured lines correspond to the same baryonic models for which we have calculated the shear power spectra in Fig.~\ref{fig:Cofl}.

The small error bars of the mock data combined with the large spread of model predictions in Fig.~\ref{fig:xrayfraction} already indicate that X-ray observations are a powerful tool to constrain baryonic parameters. In the next section we will investigate how much these additional constraints affect the cosmological bounds.


\section{Cosmological parameter inference for $\Lambda$CDM}\label{sec:likelihoodanalysis}
Based on the mock observations of the cosmic shear power spectrum (Fig.~\ref{fig:Cofl}) and the X-ray gas fraction (Fig.~\ref{fig:xrayfraction}), we now perform a likelihood analysis to quantify the constraining power of a Euclid-like survey. In this section we investigate the standard case of a $\Lambda$CDM cosmology with massive neutrinos. Different extensions of the $\Lambda$CDM model will be discussed in Sec.~\ref{sec:extensions}.

We simultaneously vary 6 cosmological ($\Omega_b$, $\Omega_m$, $\sigma_8$, $n_s$, $h_0$, and $\Sigma m_{\nu}$), 1 intrinsic-alignment ($A_{\rm IA}$), and 3 baryonic parameters ($M_c$, $\mu$, $\theta_{\rm ej}$) using a Markov Chain Monte Carlo (MCMC) sampling procedure based on a multivariate Gaussian likelihood. In order to investigate the effects of baryons on cosmology, we investigate four different scenarios that are introduced in the following:
\begin{itemize}
\item[(i)]\emph{WL only}: The likelihood sampling is performed using only mock data from the tomographic shear power spectrum (presented in Sec.~\ref{sec:WLmock}) assuming flat, uninformative priors. The baryonic parameters are allowed to vary freely within their prior-ranges. All resulting posterior contours are shown in red in the following figures.
\item[(ii)]\emph{WL+X-ray}: Additionally to the mock tomographic shear power spectrum, the mock X-ray gas fractions (presented in Sec.~\ref{sec:Xraymock}) are also included into the likelihood analysis. The baryonic parameters are still allowed to vary freely but they will be constrained by the X-ray data. The priors remain flat and uninformative. All resulting posterior contours are shown in purple.
\item[(iii)]\emph{WL+X-ray+CMB-p5}: The likelihood analysis again includes both the shear power spectrum and the X-ray gas fractions. Unlike before, however, we now assume Gaussian priors from the CMB (corresponding to the Planck 2018 results \citep{Aghanim:2018eyx}). While this is not quite the same than including CMB observations into the data vector, it allows us to estimate the constraining power obtained when combining weak-lensing with CMB observables. Note that we only use CMB priors for the five standard cosmological parameters ($\Omega_b$, $\Omega_m$, $\sigma_8$, $n_s$, $h_0$) but not for the sum of the neutrino masses ($\Sigma m_{\nu}$). All resulting posteriors of this scenario are shown in green.
\item[(iv)]\emph{WL only (no baryons)}: All baryonic effects are completely ignored during the likelihood sampling (while they are included in the mock data). This allows us to estimate the biases appearing if the prediction pipeline does not include baryonic feedback effects. We only include the tomographic shear power spectrum as mock data and we assume flat, uninformative priors for all parameters. The resulting posteriors are shown in black.
\end{itemize}
Next to these four scenarios, we also consider a fifth case where all baryonic effects are fixed to their true values (i.e. the values assumed for the mock data set). This approach, which corresponds to the idealistic situation where baryonic effects are fully understood, has already been studied in Paper I \citep{Schneider:2019snl}. In Appendix~\ref{app:fixedbaryons} we discuss it in a more general context of $\Lambda$CDM with massive neutrinos and cosmologies beyond $\Lambda$CDM.

\begin{table}[tbp]
\centering
\small
\begin{tabular}{c  c c  c }
Parameter & Mock value & Prior (\emph{WL only}, \emph{WL+X-ray}) & Prior (\emph{WL+X-ray+CMB-p5})\\
\\ [-2.5ex]
\hline
\hline
\\ [-2.5ex]
$\Omega_b$ & 0.049 & [0.04, 0.06] & 0.0010\\
$\Omega_m$ & 0.315 & [0.15, 0.42] & 0.0084\\
$\sigma_8$ & 0.786 & [0.66,0.90] & 0.0073\\
$n_s$ & 0.966 & [0.9, 1.0] & 0.0044 \\
$h_0$ & 0.673 & [0.6, 0.9] & 0.0060 \\
$\Sigma m_{\nu}$ & 0.1 & [0.0, 0.8] & [0.0, 0.8] \\
$A_{\rm IA}$ & 1.0 & [0.0, 2.0] & [0.0, 2.0]\\
$\log M_c$ & 13.8 & [13.0, 16.0] & [13.0, 16.0] \\
$\mu$ & 0.21 & [0.1, 0.7] & [0.1, 0.7] \\
$\theta_{\rm ej}$ & 4 & [2.0, 8.0] & [2.0, 8.0]\\
\\ [-2.5ex]
\hline
\\ [-2.5ex]
$10^9A_s$    & 2.025 & [1.5, 2.4] & 0.034\\
$w_{0}$ & -1.0 & [-1.5, -0.5]& [-1.5, -0.5]\\
$w_{a}$ & 0.0 & [-1.0, 1.0] & [-1.0, 1.0]\\
\\ [-2.5ex]
\hline
\\ [-2.5ex]
$\log |f_{\rm R0}|$ & -7.0 & [-7.0, -4.0] & [-7.0, -4.0] \\
\\ [-2.5ex]
\hline
\\ [-2.5ex]
$f_{\rm wdm}$ & 0.0 & [0.0, 1.0] & [0.0, 1.0]\\
$\log m_{\rm wdm}$ & $\infty$ & [-2.0, 0.0] & [-2.0, 0.0]\\

\end{tabular}
\caption{Parameter values and priors for the likelihood sampling of the \emph{WL only}, \emph{WL+X-ray}, and \emph{WL+X-ray+CMB-p5} scenarios. All priors with double values in square brackets [minimum value, maximum value] are assumed to be flat, while the ones with number only are Gaussian (the number referring to the standard deviation). For the cases where baryons are ignored or fixed, we assume the same priors than for the \emph{WL only} case. Note that for the scenarios beyond $\Lambda$CDM, $\sigma_8$ is replaced by the scalar amplitude $A_s$.}
\label{tab:prior0}
\end{table}

All free parameters and prior-ranges used for the likelihood samplings are summarised in Table~\ref{tab:prior0}. The flat priors of the scenarios (i), (ii), and (iv) are given in the third column, the Gaussian priors used in scenario (iii) are given in the fourth column. Note that the flat cosmological priors are selected to be wide enough to not affect the posteriors. One notable exception is the baryon-abundance ($\Omega_b$), which is not very sensitive to the weak-lensing signal and has been set to a range consistent with results from nucleosynthesis \citep{Steigman:2007xt}. The priors of the baryonic parameters are set to comfortably include all known predictions of hydrodynamical simulations regarding the matter power spectrum. This becomes obvious by looking at Fig.~1 of Paper I, where the grey region shows the spread in the matter power spectrum when the 3 baryonic parameters are varied within their prior-ranges.

All Markov Chain Monte Carlo (MCMC) runs are performed using the sampler {\tt UHAMMER} \citep{Akeret:2012aaa} that is build upon the code {\tt emcee} \citep{Foreman-Mackey:2013aaa}. The data vector of the tomographic shear power spectrum includes all three auto and three cross correlation spectra as shown in Fig.~\ref{fig:Cofl}. The X-ray mock observations contain both the gas fractions $f(M_{500})$ and $f(M_{2500})$ at the same three redshift bins (see Fig.~\ref{fig:xrayfraction}). Note that the gas fractions at $M_{500}$ and $M_{2500}$ are selected from different parts of the sky, i.e. every cluster is only ever used to measure either $f(M_{500})$ or $f(M_{2500})$ but never both. While this approach leads to some loss of statistical power, it allows to neglect non-diagonal contributions to the covariance matrix of the X-ray data, considerably simplifying the analysis.

\begin{figure}[tbp]
\centering
\includegraphics[width=0.49\textwidth,trim=0.2cm 0.1cm 1.2cm 0.4cm,clip]{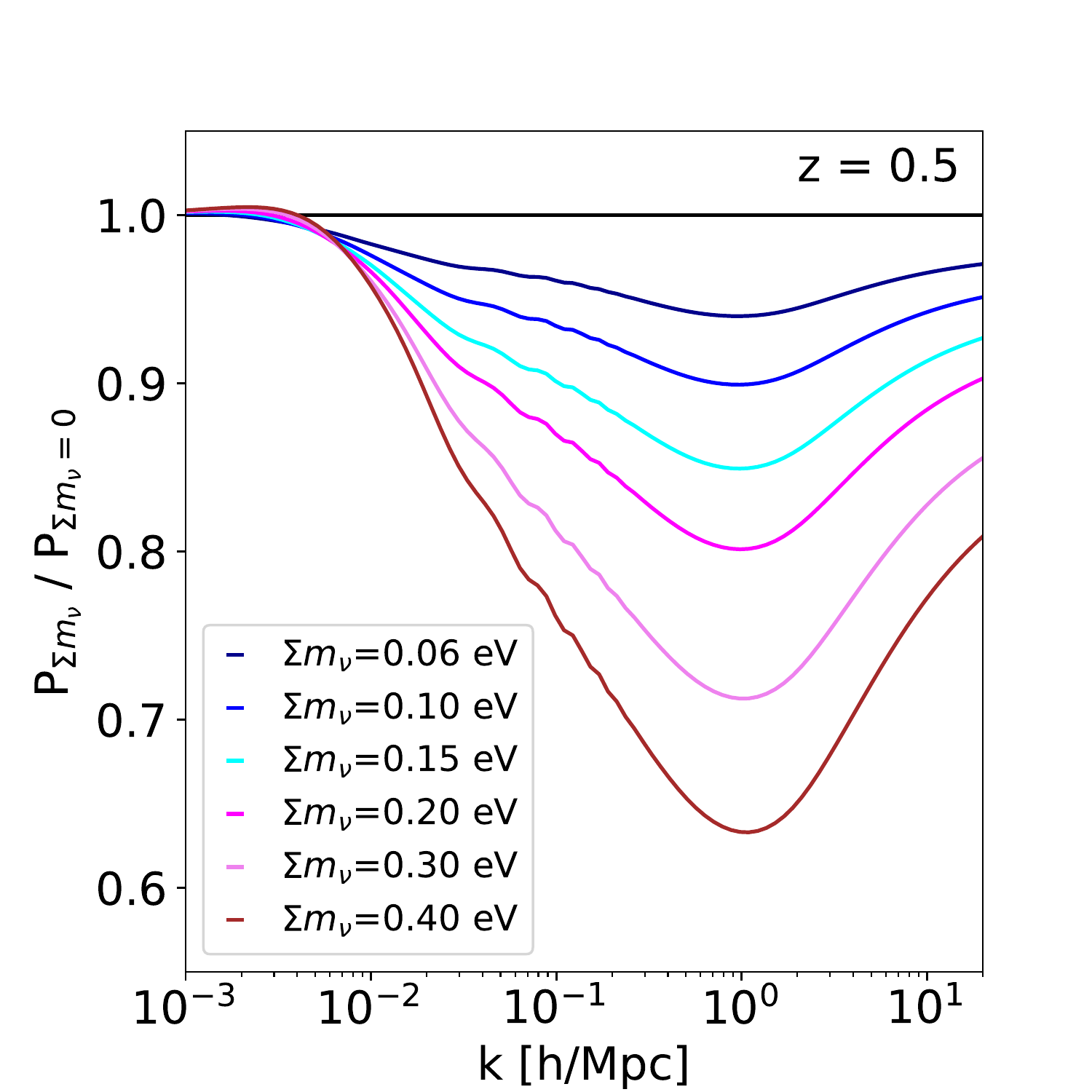}
\includegraphics[width=0.49\textwidth,trim=0.2cm 0.1cm 1.2cm 0.4cm,clip]{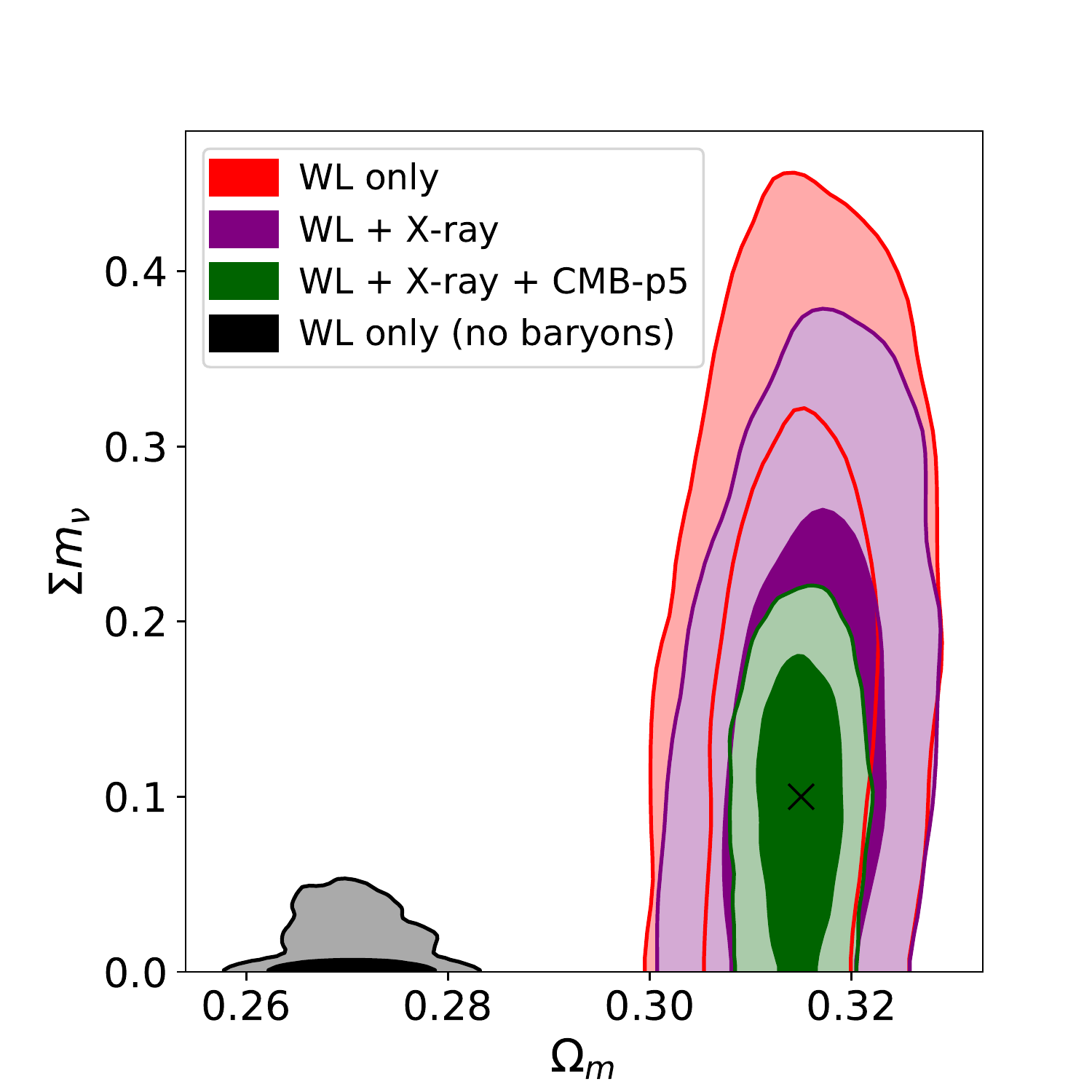}
\caption{Power spectra and likelihood contours of a $\Lambda$CDM cosmology with massive neutrinos. \emph{Left:} Ratio of selected power spectra for different values of the sum of the neutrino mass $\Sigma m_{\nu}$ divided by the $\Lambda$CDM case with massless neutrinos. \emph{Right:} Parameter contours of $\Omega_m$ and $\Sigma m_{\nu}$ for the weak-lensing only (red) and the weak-lensing plus X-ray scenario (purple) with flat, uninformative priors. The WL+Xray case with {\tt Planck} \citep{Aghanim:2018eyx} CMB priors on five parameters ($n_s$, $h_0$, $\sigma_8$, $\Omega_b$ and $\Omega_m$) is shown in green. The black contours correspond to the scenario where baryonic effects are ignored in the prediction pipeline. The cosmology assumed in the mock is indicated by the black cross.}
\label{fig:contourneutrino}
\end{figure}

Before providing a more general view on the posterior contours of cosmological and baryonic parameters, we focus on the massive neutrino parameter. The left-hand panel of Fig.~\ref{fig:contourneutrino} shows the power spectrum for different values of $\Sigma m_{\nu}$ relative to the case of massless neutrinos ($\Sigma m_{\nu}=0$ eV). The shape of the neutrino-induced power suppression is qualitatively similar to the spoon-like suppression signal from baryonic effects (see e.g. Fig. 1 of Paper I), which could lead to degeneracies between the corresponding parameters\footnote{Note, however, that despite the obvious resemblance the neutrino suppression starts at significantly larger scales \citep[see e.g. Ref.][for a more detailed discussion]{Parimbelli:2018yzv}.}. As a consequence, this would mean that combining cosmic shear with X-ray data (thereby strongly reducing the freedom from baryonic parameters) would considerably strengthen the constraints.

The right-hand panel of Fig.~\ref{fig:contourneutrino} shows the corresponding limits on the sum of the neutrino masses ($\Sigma m_{\nu}$) as a function of the matter abundance ($\Omega_m$). All other cosmological parameters are marginalised over. The cosmology of the mock (with $\Sigma m_{\nu}=0.1$ eV and $\Omega_m=0.315$) is indicated by the black cross. The red contours correspond to the forecast including weak-lensing only (scenario i). The purple contours show the combined forecast analysis, including both weak lensing and X-ray mock observations (scenario ii). The green contours furthermore include priors from current CMB data (scenario iii).

The constraining power of all three scenarios is not high enough to reliably exclude the (unrealistic) case of vanishing neutrino masses. Note that similar conclusions can be drawn from recent results of Refs.~\citep{Parimbelli:2018yzv,Copeland:2019bho}. Marginalising over $\Omega_m$ in the right-hand panel of Fig.~\ref{fig:contourneutrino} leads to an uncertainty on the sum of the neutrino mass of $\Delta \Sigma m_{\nu}=0.18$ eV (0.30 eV) for the WL-only scenario and $\Delta \Sigma m_{\nu}=0.13$ eV (0.24 eV) for the combined WL+X-ray case at the 68 percent (95 percent) confidence level. This corresponds to an improvement of $\sim30$ percent (20 percent). The uncertainty can be further reduced to $\Delta \Sigma m_{\nu}=0.05$ eV (0.1 eV) in the WL+X-ray+CMB-p5 scenario (i.e. when {\tt Planck} priors are assumed for all cosmological parameter except $\Sigma m_{nu}$), which corresponds to a 70 percent (66 percent) improvement with respect to the WL only case at the 68 percent (95 percent) confidence level. We conclude that adding eROSITA-like X-ray data leads to a noticeable but not major decrease of uncertainty for the neutrino masses. More important than including external constraints for baryonic effects is to constrain large-scale modes with data from the CMB.

Finally, the black contours in Fig.~\ref{fig:contourneutrino} show what happens if baryonic effects are completely ignored in the prediction pipeline (i.e. scenario iv). In this case, we would falsely predict a very low matter abundance $\Omega_m=0.27$ significantly below the true value of the mock data ($\Omega_m=0.315$). The constraints on the neutrino masses would be so strong that they become discrepant with current solar and atmospheric neutrino experiments. We have checked that for other cosmological parameter, such as $\sigma_8$ or $h_0$, the deviations from the true mock values are of comparable significance. This shows that ignoring baryonic effects is clearly not a viable approximation for stage-IV weak-lensing surveys. Similar conclusions have been drawn in Paper I.

\begin{figure}[tbp]
\centering
\includegraphics[width=0.98\textwidth,trim=0.1cm 0.1cm 0.5cm 0.4cm,clip]{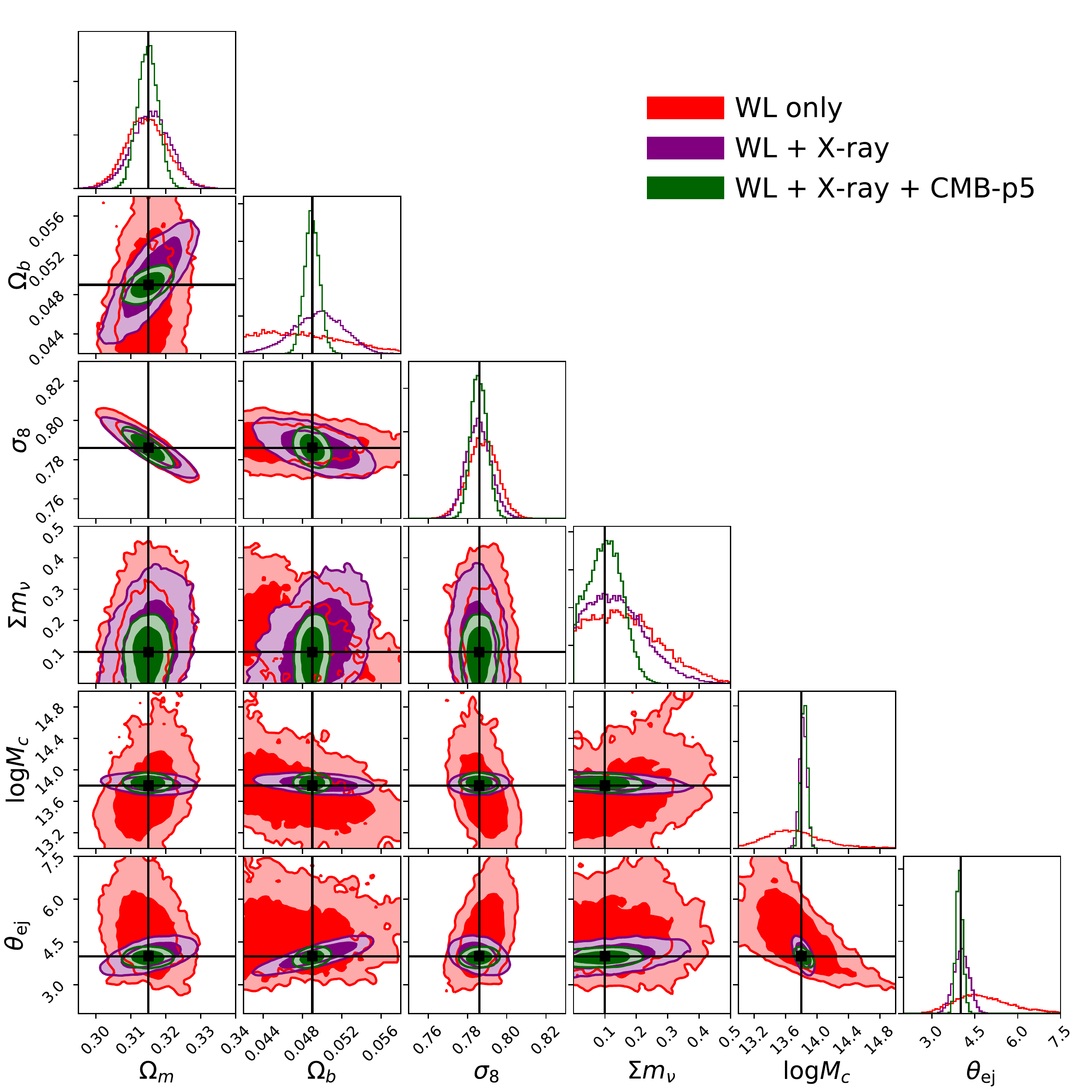}
\caption{Parameter contours of key cosmological ($\Omega_m$, $\Omega_b$, $\sigma_8$, and $m_{\nu}$) and baryonic parameters (${\log M_c}$ and $\theta_{\rm ej}$) assuming a $\Lambda$CDM model with massive neutrinos. The remaining parameters of the MCMC chain ($h_0$, $n_s$, $A_{\rm IA}$, and $\mu$) are marginalised. The WL only and the WL+X-ray results with flat, uninformative priors are shown in red and purple. The green contours correspond to the WL+X-ray scenario with {\tt Planck} \citep{Aghanim:2018eyx} CMB priors on the five parameters $n_s$, $h_0$, $\sigma_8$, $\Omega_b$ and $\Omega_m$ (but not $\Sigma m_{\nu}$). The true cosmology assumed in the mock is shown with black lines.}
\label{fig:contours_wlonly_wlxray}
\end{figure}

After specifically focusing on massive neutrinos, we now discuss other selected parameters from our forecast analysis. Fig.~\ref{fig:contours_wlonly_wlxray} illustrates the posterior contours of the cosmological parameters $\Omega_m$, $\Omega_b$, $\sigma_8$, and $\Sigma m_{\nu}$ together with the baryonic parameters ${\log M_c}$ and $\theta_{\rm ej}$. All remaining parameters ($h_0$, $n_s$, $A_{\rm IA}$, and $\mu$) are marginalised over. The red contours again show the WL-only case (scenario i), while the purple and green contours correspond to the combined WL+X-ray scenarios with flat, uninformative priors (scenario ii) and with Gaussian priors in accordance with current CMB constraints (scenario iii).

Fig.~\ref{fig:contours_wlonly_wlxray} shows that including eROSITA data of X-ray gas fractions provides stringent constraints on baryonic feedback parameters. The expected errors on both $\log M_c$ and $\theta_{\rm ej}$ are strongly reduced with respect to the weak-lensing only case. The same is true regarding the third baryonic parameter $\mu$ as we will see later on. However, Fig.~\ref{fig:contours_wlonly_wlxray} also illustrates that there are no strong degeneracies between baryonic and cosmological parameters. This means that finding better and better constraints for the baryonic parameters only leads to moderate improvements in terms of cosmology. This statement is further emphasised in Appendix~\ref{app:fixedbaryons}, where we show that fixing baryonic parameters to their true values (assumed in the mock) does only led to a moderate further decrease of the posterior contours.

As a side-note it should be emphasised that Euclid-like observations of the cosmic shear power spectrum alone make it possible to constrain the baryonic parameters beyond the original prior ranges (albeit not with the same constraining power than the X-ray data). This means that stage-IV weak-lensing surveys will not only be able to teach us more about cosmology but also about baryonic feedback processes of galaxies. This conclusion, already mentioned in Paper I, remains true for the more realistic case of a $\Lambda$CDM model with massive neutrinos.

Another conclusion that can be drawn from Fig.~\ref{fig:contours_wlonly_wlxray} is that the $\Omega_b$ parameter can be constrained with the WL+X-ray combination but remains unconstrained in the WL only scenario. The latter is not surprising as weak lensing only probes the total matter distribution which is not sensitive to the $\Omega_b$ parameter. The former is due to the fact that the X-ray gas fractions are not only sensitive to baryonic feedback but also carry some cosmological information. More explicitly, the X-ray gas fractions are sensitive to the cosmic baryon fraction $f_b=\Omega_b/\Omega_m$, which sets the overall normalisation of the data points visible in Fig.~\ref{fig:xrayfraction}.

\begin{figure}[tbp]
\centering
\includegraphics[width=0.98\textwidth,trim=2.0cm 0.4cm 2.2cm 0.4cm,clip]{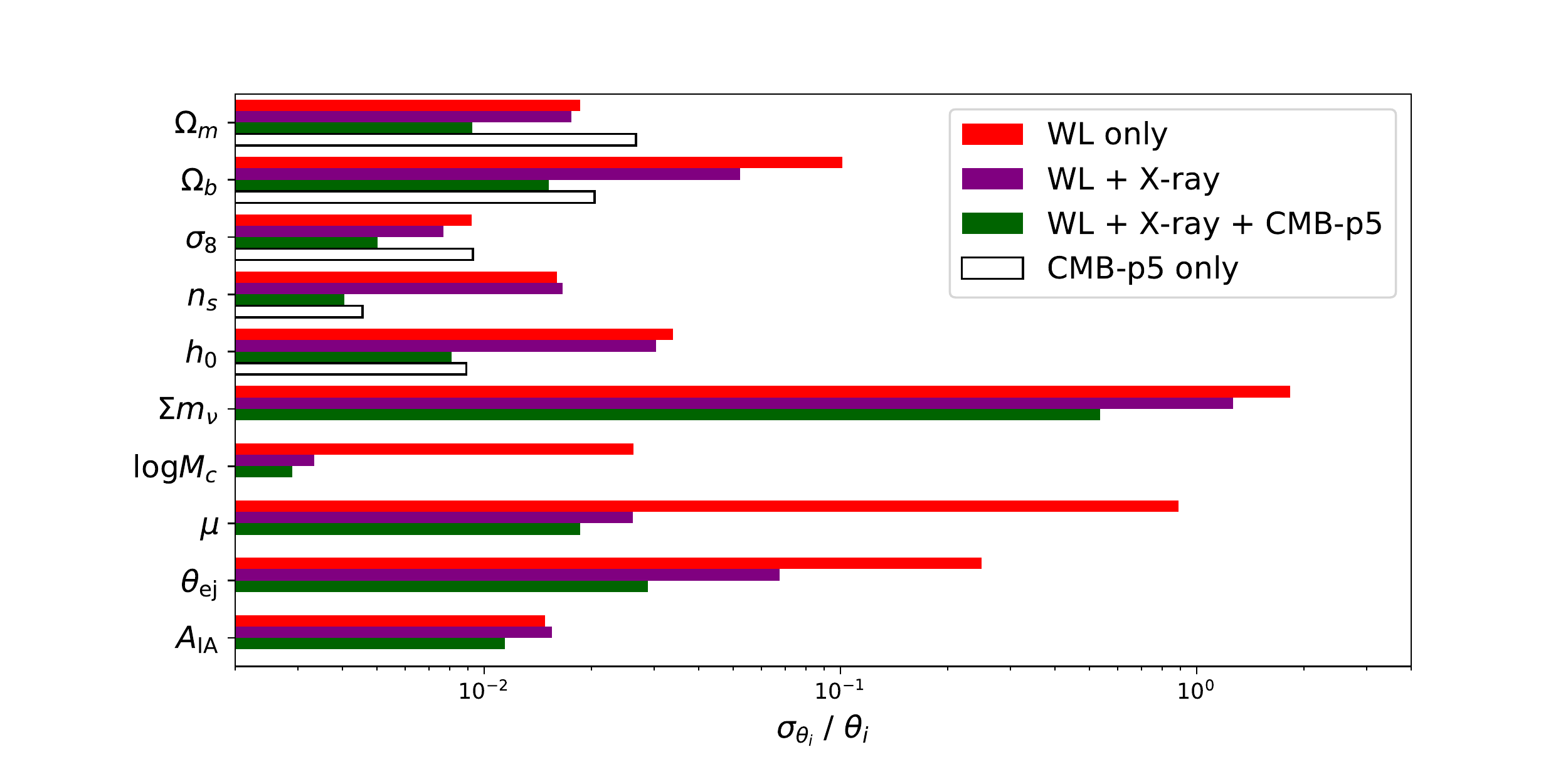}
\caption{Marginalised errors ($\sigma_{\theta_i}$) of all model parameters (at 68 percent confidence level) normalised by the corresponding default parameter value ($\theta_i$). The WL only and the WL+X-ray results with flat, uninformative priors are shown as red and purple bars. The green bars correspond to the WL+X-ray scenario with Gaussian priors on $n_s$, $h_0$, $\sigma_8$, $\Omega_b$ and $\Omega_m$ (but not $\Sigma m_{\nu}$) from the CMB experiment {\tt Planck} \citep{Aghanim:2018eyx}. The black empty bars provide the size of these CMB priors. They indicate the current state-of-the art on parameter estimates.}
\label{fig:sigma_wlonly_wlxray}
\end{figure}

A summary of the results discussed above is provided in Fig.~\ref{fig:sigma_wlonly_wlxray}, where we plot the marginalised errors $\sigma_i$ (at the 68 percent confidence level) of all free model parameters $i$ divided by their default parameter value $\theta_i$. The results from the WL only scenario (i) is again shown in red, while the WL+X-ray scenarios without and with CMB priors (ii and iii) are shown in purple and green.

Fig.~\ref{fig:sigma_wlonly_wlxray} shows that adding X-ray information to the WL data leads to a reduction of the errors on the baryonic parameters by more than $70$ percent for $\theta_{\rm ej}$, $85$ percent for $\log M_{c}$ and 95 percent for $\mu$. As already shown above, the improvement regarding the cosmological parameters is not nearly as impressive. Only the errors on $\Omega_b$ and $\Sigma m_{\nu}$ decrease by $\sim30$ percent or more, while the errors of the remaining parameters decrease by only $\sim5$-20 percent ($\Omega_m$, $\sigma_8$, and $h_0$) or not at all ($n_s$ and $A_{\rm IA}$). A much more significant improvement of all parameter uncertainties is obtained if CMB priors are included. In this case, the errors of all cosmological parameters shrink by $\sim 50$ percent or more. This also has an effect on the errors of the baryonic parameters which become even tighter than for the WL+X-ray case with flat priors.

The main conclusion obtained from Fig.~\ref{fig:sigma_wlonly_wlxray} is that in terms of cosmological parameter estimates, it is more promising to further constrain large scales with CMB data than to reduce baryonic uncertainties at small scales. Note that the decrease of errors when going from purple to green (scenario ii to iii) is not uniquely driven by the CMB priors but is due to a combination of CMB and WL likelihoods. This becomes evident when comparing the error estimates from the WL+Xray+CMB-p5 scenario to the CMB prior ranges alone (green versus black empty bars) where the former are significantly tighter than the latter, especially regarding the parameters $\Omega_m$ and $\sigma_8$.

Looking at both the current CMB errors from {\tt Planck} and the expected uncertainties from WL+Xray observations also allows to compare the current state-of-the-art with future expectations. Fig.~\ref{fig:sigma_wlonly_wlxray} shows that significant improvement in terms of cosmological parameters can be expected for $\Omega_m$ and $\sigma_8$ but not for $\Omega_b$, $n_s$, and $h_0$. This is not surprising as the former two are known to be particularly sensitive parameters for weak lensing.

Finally, let us compare the results of Fig.~\ref{fig:sigma_wlonly_wlxray} to the Euclid-specific forecast analysis of \citet[henceforth B19]{Blanchard:2019oqi}. Compared to this paper, B19 used a setup that is closer to Euclid survey characteristics, resulting in a somewhat different galaxy distribution and redshift range, 10 instead of 3 redshift bins, and a survey area of 15000 deg$^2$ instead of 20000 deg$^2$. On the other hand, they relied on the less accurate Fisher matrix analysis instead of a full MCMC likelihood analysis and they did not account for neutrino mass variations nor baryonic feedback effects.

Fig.~\ref{fig:sigma_wlonly_wlxray} shows that our forecast results for $\Omega_m$ and $\sigma_8$ are very similar to the ones from B19 (i.e. their WL only case assuming a flat universe). On the other hand, we obtain smaller uncertainties for $\Omega_b$, $n_s$, and $h_0$, which is, at least partially, due to tighter prior choices for $\Omega_b$ and $n_s$ (see Table~\ref{tab:prior0} and related discussion in the text).

In the present section, we have established that the cosmic shear power spectrum alone is able to provide surprisingly tight constraints on cosmological parameters as long as baryonic effects are properly parametrised. Furthermore, we showed that it is possible to efficiently constrain the baryonic parameters with external data from X-ray gas fractions, leading to a further reduction of cosmological parameter uncertainties at the order of 10-30 percent. A much stronger improvement of order 50 percent or more is expected if the weak-lensing plus X-ray data is combined with data from the CMB.


\section{Testing cosmological extensions}\label{sec:extensions}
So far, we have limited ourselves to the standard case of a six parameter cosmological model with cosmological constant and one massive neutrino species. However, one of the main goals of future cosmological surveys is not only to pin down well-known parameters to better and better precisions, but to test possible deviations or extensions of the standard $\Lambda$CDM cosmology. In this section, we go beyond the minimal cosmological model and determine the expected constraints for three next-to-minimal cosmologies. The first case is the well known dark energy model with dynamical equation of state, parametrised by $w_0$ and $w_a$. The second case is the $f(R)$ modified gravity model of \citet{Hu:2007nk} that has become the best studied toy-model for modified gravity extensions with screening mechanism. The third model consists of a mixed dark matter cosmology, where the dark matter sector is described by a perfectly cold and a warm (or hot) particle component. All three models exhibit modifications of the power spectrum that become larger towards higher wave modes. This could potentially lead to degeneracies with the baryonic suppression signal, making them ideal targets of investigation.

Following the procedure set up in the beginning of Sec.~\ref{sec:likelihoodanalysis}, we investigate four different cases, the {WL only} (i), the {WL+X-ray} (ii) , the {WL+X-ray+CMB-p5} (iii), and the {WL only (no baryons)} scenario (iv), performing a likelihood analysis for each one of them. A fifth {WL only} scenario with fixed baryonic parameters is presented in Appendix~\ref{app:fixedbaryons}. Regarding scenario (iii), note that the CMB priors from {\tt Planck} only apply to the five standard cosmological parameters  ($n_s$, $h_0$, $A_s$, $\Omega_b$ and $\Omega_m$) but neither to the sum of the neutrino masses ($\Sigma m_{\nu}$) nor the parameters describing the $\Lambda$CDM extension\footnote{For all beyond-$\Lambda$CDM scenarios discussed in this sections, we replace the cosmological parameter $\sigma_8$ by the scalar amplitude $A_s$.}.  

In the following sections, we are going to focus solely on the key parameters of the wCDM, fRCDM, and $\Lambda$MDM models without showing more complete representations of the posteriors. However, the interested reader may be referred to Appendix \ref{app:extendedcosmologies} for a more detailed analysis of the expected uncertainties on all remaining parameters of the MCMC sampling procedure.

\subsection{Dynamical dark energy (wCDM)}
The wCDM model is one of the best studied extensions of the $\Lambda$CDM standard cosmological paradigm. It is based on the assumption that the dark energy component is dynamical (instead of a constant $\Lambda$) with an equation-of-state $w_{\rm de}(z)$. A common and well tested parametrisation is \citep{Chevallier:2000qy,Linder:2002et}
\begin{equation}\label{DEEOS}
w_{\rm de}(z) = w_0 + w_a\frac{z}{1+z}.
\end{equation}
This equation reduces to the standard $\Lambda$CDM case for $w_{de}(z)=-1$ (which means that $w_0=-1$ and $w_a=0$), whereas $w_{\rm de}>-1$ resembles standard quintessence models \citep{Scherrer:2015tra} while $w_{\rm de}<-1$ is known as phantom dark energy \citep[][]{Ludwick:2017tox}. Following e.g. Ref.~\citep{Abbott:2018xao}, we impose $w_0+w_a\leq0$ as a hard prior. The other priors on the individual parameters are given in Table~\ref{tab:prior0}.

In order to estimate the expected constraints on dynamical dark energy from a Euclid-like survey, we perform likelihood samplings including all baryonic, intrinsic-alignment, and cosmological parameters from before, plus the new parameters $w_0$ and $w_a$ from Eq.~(\ref{DEEOS}). We rely on the same prediction pipeline for the matter power spectrum, i.e. the {\tt revised halofit} model of \citet{Takahashi:2012em}. The mock data set is also the same than the one presented above, corresponding to the $\Lambda$CDM case with $w_0=-1$ and $w_a=0$. The goal is to estimate how well we will be able to constrain deviations from $\Lambda$CDM in the $w_0-w_a$ plane with the weak-lensing shear spectrum alone and by combining it with the X-ray data set from eROSITA. Furthermore, we want to quantify the bias that is introduced if baryonic effects are ignored.

\begin{figure}[tbp]
\centering
\includegraphics[width=0.49\textwidth,trim=0.2cm 0.1cm 1.2cm 0.4cm,clip]{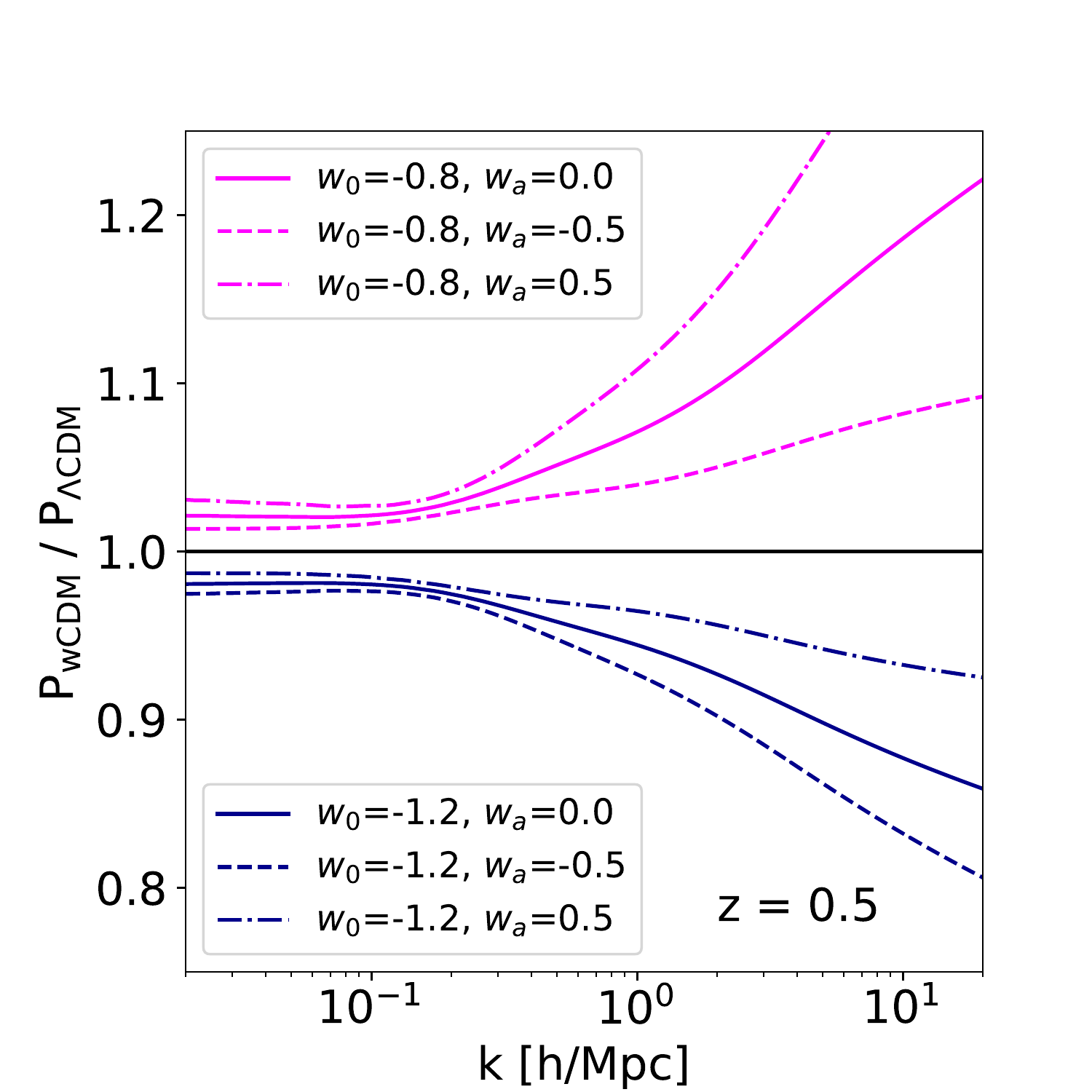}
\includegraphics[width=0.49\textwidth,trim=0.2cm 0.1cm 1.2cm 0.4cm,clip]{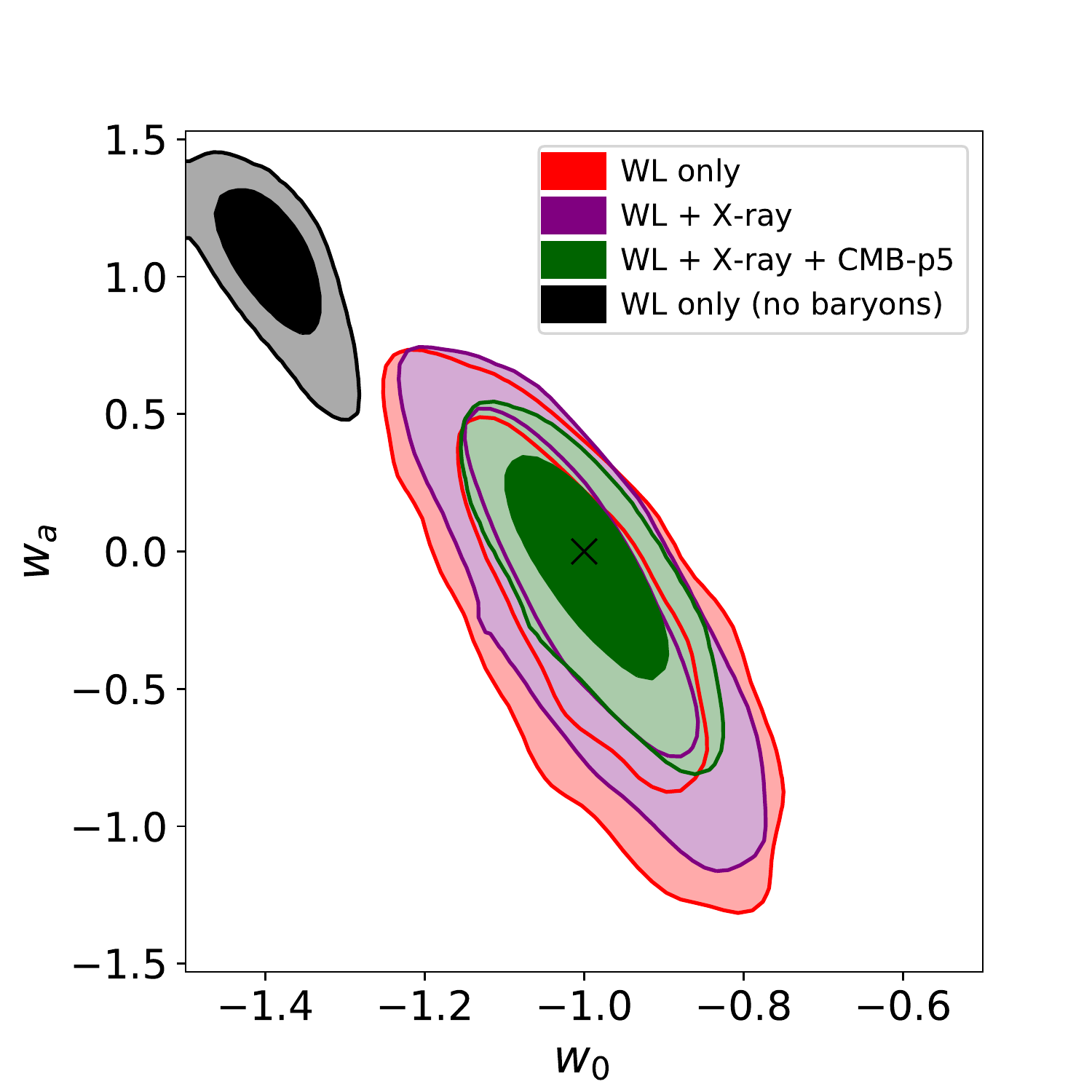}
\caption{Power spectra and likelihood contours of a wCDM cosmology with dynamical dark energy equation of state. \emph{Left:} Ratio of selected power spectra for different values of $w_0$ and $w_a$ with respect to the standard $\Lambda$CDM case ($w_0=-1$, $w_a=0$). \emph{Right:} Comparison of the $w_0$ and $w_a$ posterior contours for different forecast scenarios, where all other parameters are marginalised over. The WL only and the WL+X-ray cases with flat, uninformative priors are shown in red and purple. The green contours correspond to the WL+X-ray scenario with CMB {\tt Planck} \citep{Aghanim:2018eyx} priors on the five standard parameters ($n_s$, $h_0$, $A_s$, $\Omega_b$, and $\Omega_m$). The black contours indicate what happens if baryonic effects are ignored. The parameters of the mock are shown as black cross.}
\label{fig:contourW0WA}
\end{figure}

The left-hand panel of Fig.~\ref{fig:contourW0WA} illustrates the relative effect on the matter power spectrum at z=0.5 for a dark energy equation-of-state deviating from the $\Lambda$CDM case. We show two example-cases for $w_0$, where the solid lines correspond to $w_a=0$, while the dashed and dash-dotted lines give the results for $w_a=-0.5$ and 0.5. In general, the amplitude of the power spectrum becomes larger (smaller) for larger (smaller) values of $w_0$ and $w_a$. Furthermore, the difference between the wCDM and the $\Lambda$CDM case increases towards smaller physical scales (larger $k$-modes).

The right-hand panel of Fig.~\ref{fig:contourW0WA} shows the results of the likelihood analysis in terms of the $w_0-w_a$ posterior contours while all other parameters are marginalised over. The true cosmology of the mock data set is indicated as a black cross. Following the same colour scheme than in the previous section, the results from the WL only scenario (i) are shown in red, while the WL+X-ray scenario with flat, uninformative priors (ii) is shown in purple. The green contours correspond to the WL+X-ray scenario with {\tt Planck} \citep{Aghanim:2018eyx} priors on the five standard parameters ($n_s$, $h_0$, $A_s$, $\Omega_b$, and $\Omega_m$), i.e. case (iii). Note again that the remaining cosmological parameters ($\Sigma m_{\nu}$, $w_0$, and $w_a$) do not include CMB prior information.

Fig.~\ref{fig:contourW0WA} shows that both additional X-ray data and CMB priors lead to a noticeable reduction of the dark energy parameter contours. The marginalised uncertainty on $w_0$ decreases from $\Delta {w_0}=0.101$ (0.200), to $\Delta {w_0}=0.092$ (0.177), and $\Delta {w_0}=0.062$ (0.120) for the scenarios (i)-(iii) at the 68 percent (95 percent) confidence level. The uncertainty on $w_a$, on the other hand, goes from $\Delta {w_a}=0.443$ (0.838), to $\Delta {w_a}=0.408$ (0.752), and $\Delta {w_a}=0.256$ (0.502) at the 68 percent (95 percent) confidence level. The rather moderate differences between scenario (i) and (ii) indicate that current uncertainties related to baryonic feedback effects do not significantly degrade the expected errors on the dynamical dark energy parameters. This conclusion is confirmed in Appendix~\ref{app:fixedbaryons}, where we show that the idealistic scenario of fixed baryonic parameters does not lead to tighter constraints either. Including CMB priors, on the other hand, helps to reduce the errors on $w_0$ and $w_a$ at a more significant level, highlighting the advantage of a combined approach for cosmological parameter inference.

The combined accuracy of the dark energy equation-of-state parameters can be quantified using the Figure of Merit (FoM), which we define as 
\begin{equation}
{\rm FoM} = (\Delta {w_0}\times\Delta{w_a})^{-1},
\end{equation}
where $\Delta {w_0}$ and $\Delta{w_a}$ refer to the errors at the 68 percent confidence level. We obtain a Figure of Merit of ${\rm FoM}=22$ for the WL only, ${\rm FoM}=27$ for the WL+X-ray, and ${\rm FoM}=63$ for the WL+X-ray+CMB-p5 scenarios. The idealistic case of fixed baryonic parameters (but without CMB priors) leads to ${\rm FoM}=32$ (see Appendix~\ref{app:fixedbaryons}). It is worth noticing that these results are in reasonable agreement with the Euclid-specific forecast paper of B19 \citep{Blanchard:2019oqi}.

Finally, we investigate what happens if baryonic effects are ignored in the prediction pipeline (i.e. scenario iv). In Fig.~\ref{fig:contourW0WA}, the corresponding contours are given in black. They show a strong preference for a phantom dark energy model with biases on $w_0$ and $w_a$ of more than 5 standard deviations. This means that ignoring baryons would lead to a false exclusion of a cosmological constant as the driver behind the accelerated expansion of the universe. It is therefore evident that including baryonic effects into the WL analysis pipeline will be essential for studying dark energy extensions with stage-IV lensing surveys.

\subsection{Chameleon $f(R)$ gravity}
The $f(R)$ gravity model (fRCDM) is characterised by a generalisation of the Einstein-Hilbert action where the Ricci scalar $R$ is replaced with an arbitrary function $f(R)$. One of the best studied cases is the \citet{Hu:2007nk} parametrisation of $f(R)$, which results in a screening mechanism where noticeable deviation from general relativity only manifest themselves in low-density regions, thereby evading the most stringent constraints from solar-system tests. The functional form contains two free parameters $f_{\rm R0}$ and $n$ \citep[see Eqs.~3 and 30 of Ref.][]{Hu:2007nk}, here we restrict ourselves to the case with fixed $n=1$ and varying $f_{\rm R0}$.

We use the fitting function of \citet{Winther:2019mus} to quantify the effect of $f(R)$ gravity on the nonlinear matter power spectrum. The fit is given by
\begin{equation}\label{fRfit}
\frac{P_{\rm fRCDM}}{P_{\rm \Lambda CDM}}= 1+B\times\left[\frac{1+Ck}{1+Dk}\right]\times\arctan\left[Ek\right]\times\exp\left[F+Gk\right],
\end{equation}
where $B$, $C$, $D$, $E$, $F$, and $G$ are second-order polynomial functions of redshift with each of the polynomial coefficients being themselves second order polynomial functions. The best-fitting coefficients are given in Table II of \citet{Winther:2019mus}.

\begin{figure}[tbp]
\centering
\includegraphics[width=0.49\textwidth,trim=0.2cm 0.1cm 1.2cm 0.4cm,clip]{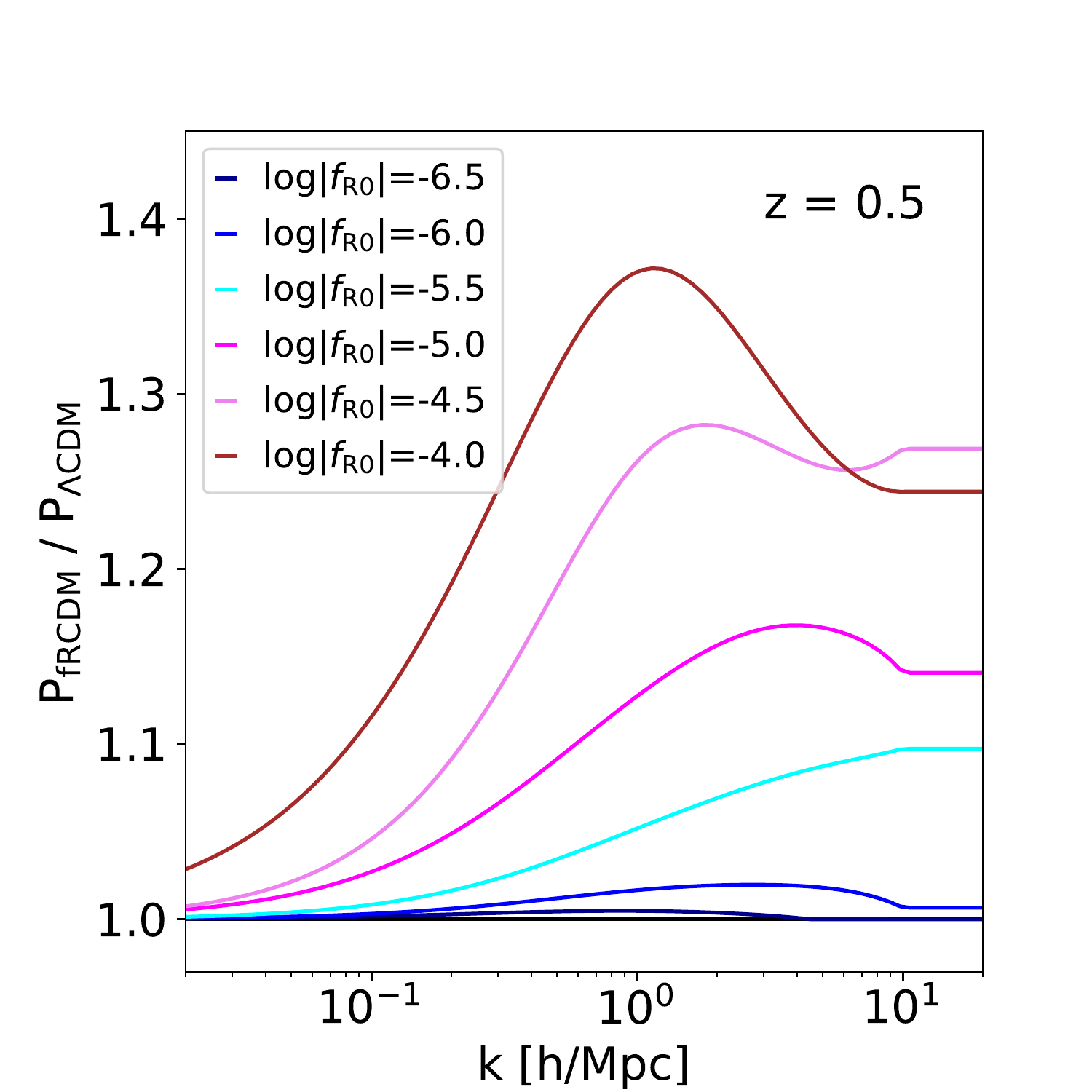}
\includegraphics[width=0.49\textwidth,trim=0.2cm 0.1cm 1.2cm 0.4cm,clip]{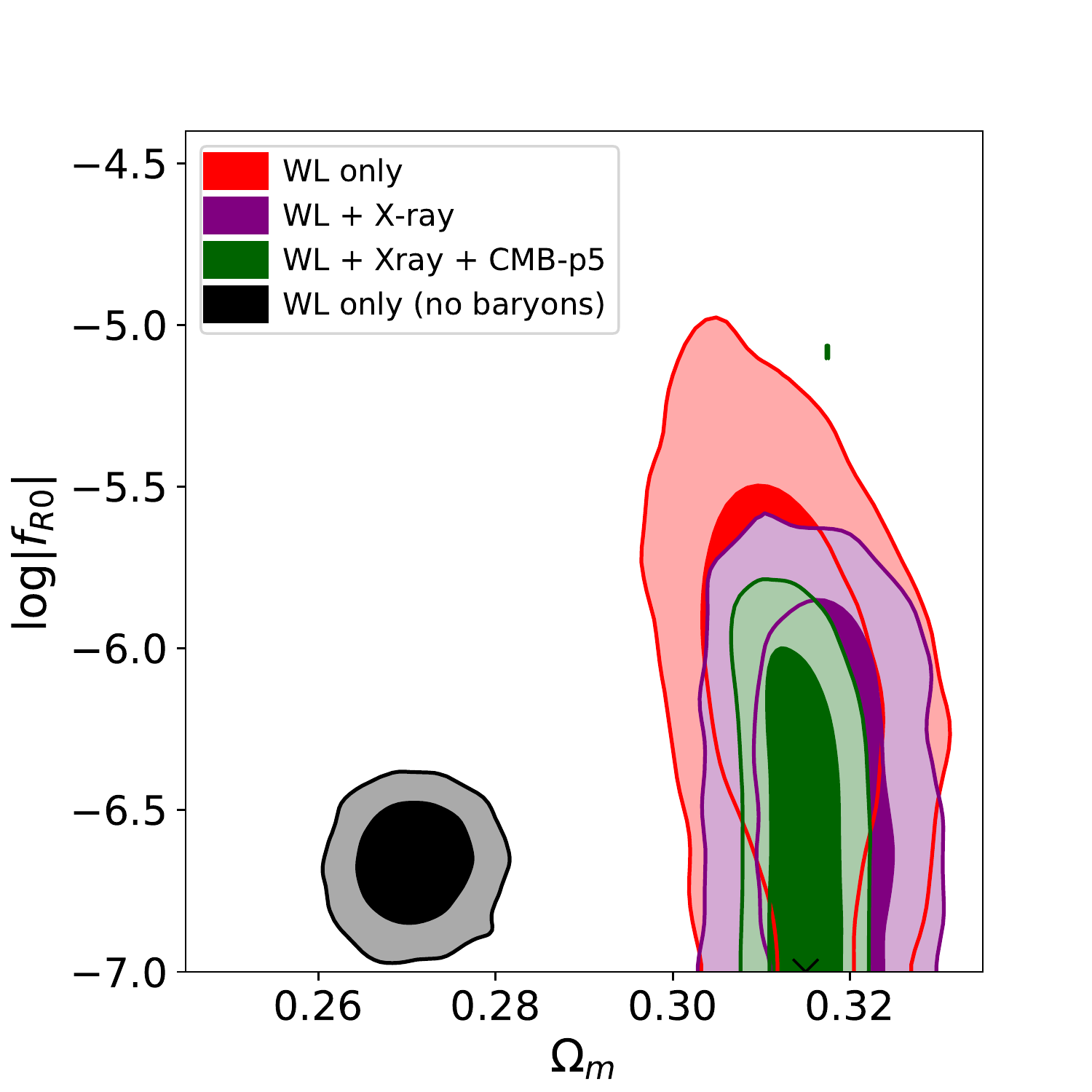}
\caption{Power spectra and likelihood contours of a $f(R)$ modified gravity scenario (fRMDM), where the $f_{\rm R0}$ parameter controls the deviation from general relativity. \emph{Left:} Ratio of selected power spectra for different values of $\log |f_{\rm R0}|$. Below $\log |f_{\rm R0}|\sim-7$, the power spectrum becomes indistinguishable to $\Lambda$CDM. \emph{Right:} Comparison of the $\Omega_m$ and $\log |f_{\rm R}|$ posterior contours for different forecast scenarios, where all other parameters are marginalised over. The WL only and the WL+X-ray cases with flat, uninformative priors are shown in red and purple. The green contours correspond to the WL+X-ray scenario with CMB {\tt Planck} priors on the five standard parameters ($n_s$, $h_0$, $\sigma_8$, $\Omega_b$ and $\Omega_m$). The black contours indicate what happens if baryonic effects are ignored. The parameters of the mock are shown as black cross.}
\label{fig:contourfR}
\end{figure}

The relative power spectrum of fRCDM obtained with Eq.~(\ref{fRfit}) at $z=0.5$ is shown in the left-hand panel of Fig.~\ref{fig:contourfR}. It is characterised by an excess of power at small scales ($k\gtrsim0.1$ h/Mpc) with respect to the standard $\Lambda$CDM case. The amplitude of the excess depends on the $f_{\rm R0}$ parameter. Below $|f_{\rm R0}|\sim 10^{-7}$, the results are indistinguishable from $\Lambda$CDM. Note that all relative power spectra become flat at $k>10$ h/Mpc. This is not a physical effect but corresponds to the maximum scale to which the fitting function has been calibrated.

We model the $f(R)$ power spectrum by simply multiplying Eq.~(\ref{fRfit}) to the nonlinear power spectrum from the {\tt revised halofit} model \citep{Takahashi:2012em}. This approach is justified by Ref.~\citep{Winther:2019mus}, where it is shown that Eq.~(\ref{fRfit}) is fairly insensitive to modifications of cosmological parameters. The MCMC sampling is performed using the same parameters as above. This means that, next to the new model parameter $f_{\rm R0}$, we vary all cosmological $\Lambda$CDM parameters (including the sum of the neutrino mass) as well as the intrinsic-alignment and the three baryonic parameters.

The resulting posterior contours of $\Omega_m$ and $|f_{\rm R0}|$ are shown in the right-hand panel of Fig.~\ref{fig:contourfR}. All other parameters are marginalised over. As before, the red and purple contours correspond to the WL only and the WL+X-ray scenarios with flat, uninformative priors. The WL+X-ray scenario with {\tt Planck} priors on the five standard parameters ($n_s$, $h_0$, $\sigma_8$, $\Omega_b$ and $\Omega_m$) is shown in green. The other cosmological parameters ($\Sigma m_{\nu}$ and $f_{\rm R0}$) do not include prior information from the CMB.

Marginalising over $\Omega_m$ in Fig.~\ref{fig:contourfR} results in constraints of $\log |f_{\rm R0}|<-5.7$ ($-5.3$) for the WL only and $\log |f_{\rm R0}|<-6.0$ ($-5.8$) for the WL+X-ray scenarios at the 68 percent (95 percent) confidence level. The substantial improvement can be explained by the fact that baryonic feedback and $f(R)$ gravity both lead to modifications of the power spectrum that are similar in shape and amplitude. Fixing baryonic effects with X-ray data therefore allows to put stronger limits on $|f_{\rm R0}|$. Adding CMB priors on the five standard cosmological parameters, on the other hand, leads to $\log |f_{\rm R0}|<-6.1$ ($-5.9$) at the 68 percent (95 percent) confidence level, which is only a mild further improvement compared to the case without CMB priors. In Appendix~\ref{app:fixedbaryons} we show that the same is true for the WL only scenario where baryonic effects are fixed. We therefore conclude that combining WL with cluster gas fractions from eROSITA allows us to obtain optimal constraints on $f(R)$ gravity. Note that the expected limit on $|f_{\rm R0}|$ is significantly tighter than current constraints from cosmological probes \citep{Jain:2013wgs,Bel:2014awa,Lombriser:2014dua,Peirone:2016wca}.

Finally, we again run a forecast analysis for the case where baryonic effects are ignored in the prediction pipeline. The results are given as black contours in Fig.~\ref{fig:contourfR}. They show preference for a $f(R)$ gravity model with $\log |f_{\rm R0}|\sim-6.7$ compared to the standard $\Lambda$CDM case. However, and more importantly, the black contours show that ignoring baryons results in a limit $\log |f_{\rm R0}|<6.5$ at the 95 percent confidence level, which is overly optimistic with respect to the realistic limits presented above.

\subsection{Mixed dark matter ($\Lambda$MDM)}
The $\Lambda$MDM model is based on a dark matter sector with a cold and a warm (or hot) dark component. There are multiple particle physics configurations where such a scenario could arise. For example, dark matter could be made of sterile neutrinos, where the warm DM component consists of particles produced via the standard \citet[DW,][]{Dodelson:1993je} mechanism, while the cold component could arise due to the decay of a scalar singlet \citep{Merle:2014xpa,Abazajian:2019ejt}. Another option is multiple axion DM with an ultra-light axion-like particle playing the role of warm and a Peccei-Quinn axion the role of cold dark matter \citep{Marsh:2013ywa, Hui:2016ltb}.

While the details of clustering may depend on the specific particle physics model, all of these models exhibit a smoothed step-like suppression feature of the linear matter power spectrum, where the position of the step depends on the mass of the warm component and its amplitude is governed by the fraction of warm to cold species \citep[see e.g. Ref.][]{Anderhalden:2012qt}. Due to nonlinear structure formation this initial step gets transformed into a more gradual suppression, which, however, is shallower than the suppression caused by pure warm dark matter.

In this paper we focus on the case of a subdominant warm component that has become non-relativistic before the time of photon decoupling, i.e. $m_{\rm wdm}\gtrsim10$ eV \citep{Aghanim:2018eyx}. This means we do not adopt the number of relativistic species ($N_{\rm eff}$) assumed in the analysis. The correction to the nonlinear matter power spectrum is calculated following the fitting function of \citet{Kamada:2016vsc}:
\begin{equation}\label{PMDM}
\frac{P_{\rm \Lambda MDM}}{P_{\rm \Lambda CDM}}=1-A(f_{\rm wdm})+\frac{A(f_{\rm wdm})}{(1+k/k_d)^{0.7441}},\hspace{0.3cm}A(f_{\rm wdm})=1-\exp\left[-\frac{1.551\times f_{\rm wdm}^{0.576}}{1-f_{\rm wdm}^{1.263}}\right].
\end{equation}
The specific wave mode $k_d$ is given by
\begin{equation}
k_d=388.8\left(\frac{m_{\rm wdm}}{\rm 10^3\, eV}\right)^{2.207}f_{\rm wdm}^{-5/6}D(z)^{1.583},
\end{equation}
where $D(z)$ is the growth factor normalised at redshift zero. The fitting function has two free parameters, the mass of the warm relic ($m_{\rm wdm}$) in eV and the fraction of warm to total dark matter ($f_{\rm wdm}=\Omega_{\rm wdm}/\Omega_{\rm dm}$). Note that $m_{\rm wdm}$ consists of the thermal-like WDM mass term, which only coincides with the true particle mass for the case of a thermally produced dark matter relic \citep[see e.g. Refs.][]{Viel:2005qj,Merle:2015vzu}. For the non-thermally produced DW sterile neutrinos (sn), the mass conversion is $m_{\rm sn}\simeq3.9\times (m_{\rm wdm}/{\rm keV})^{1.294}$ keV \citep{Bozek:2015bdo}. For the ultra-light axion scenario, an approximate conversion can be found by equating the half-mode scales as in Ref.~\citep{Lidz:2018fqo}.

We now perform a cosmological parameter inference varying the new model parameters $w_{\rm wdm}$ and $f_{\rm wdm}$ plus all previous baryonic, intrinsic-alignment, and cosmological parameters including $\Sigma m_{\nu}$. The priors are described in Table~\ref{tab:prior0}. Regarding the mock observation, we use the same data as above, which corresponds to a $\Lambda$CDM model with perfectly cold dark matter, i.e. $f_{\rm wdm}=0$.

\begin{figure}[tbp]
\centering
\includegraphics[width=0.49\textwidth,trim=0.2cm 0.1cm 1.2cm 0.4cm,clip]{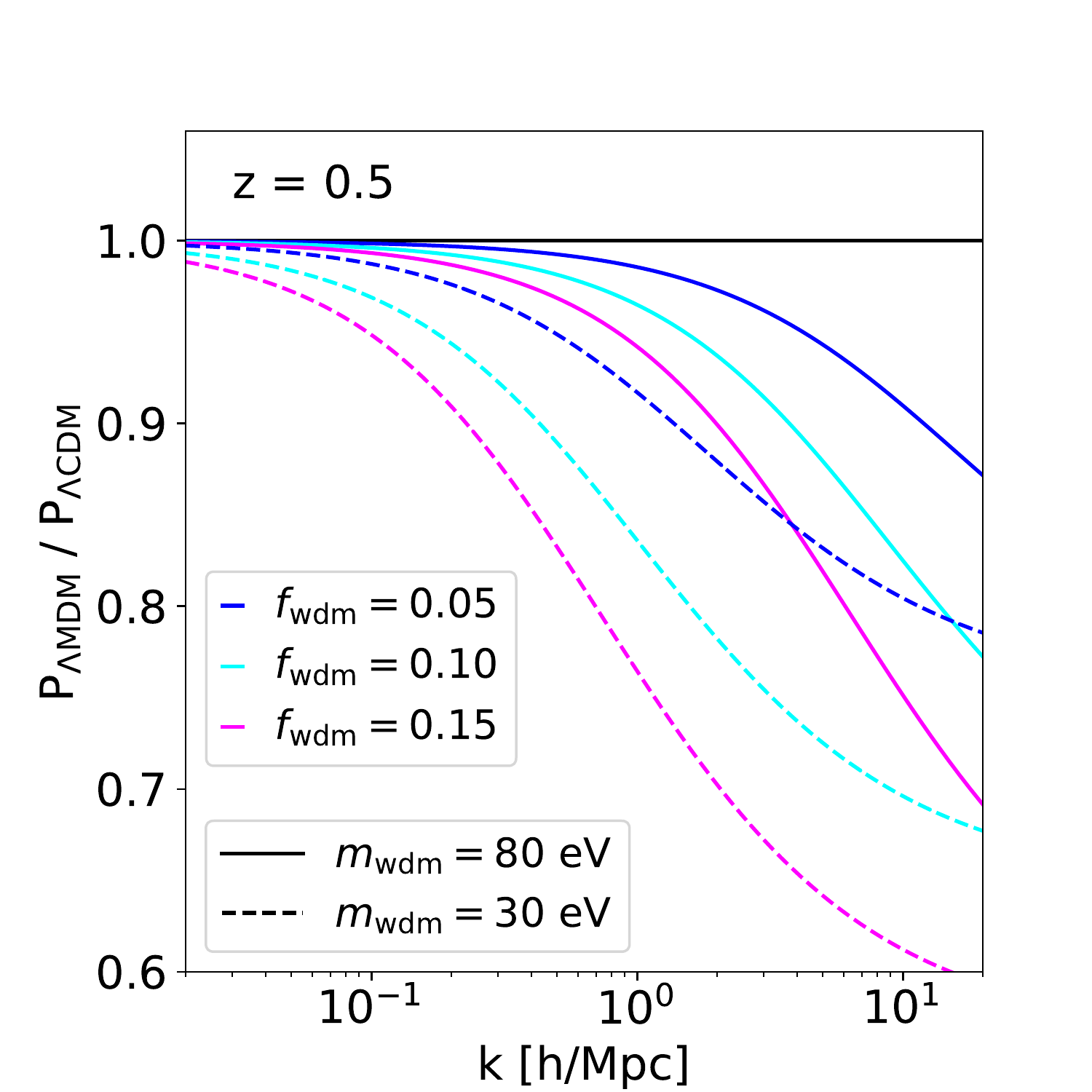}
\includegraphics[width=0.49\textwidth,trim=0.2cm 0.1cm 1.2cm 0.4cm,clip]{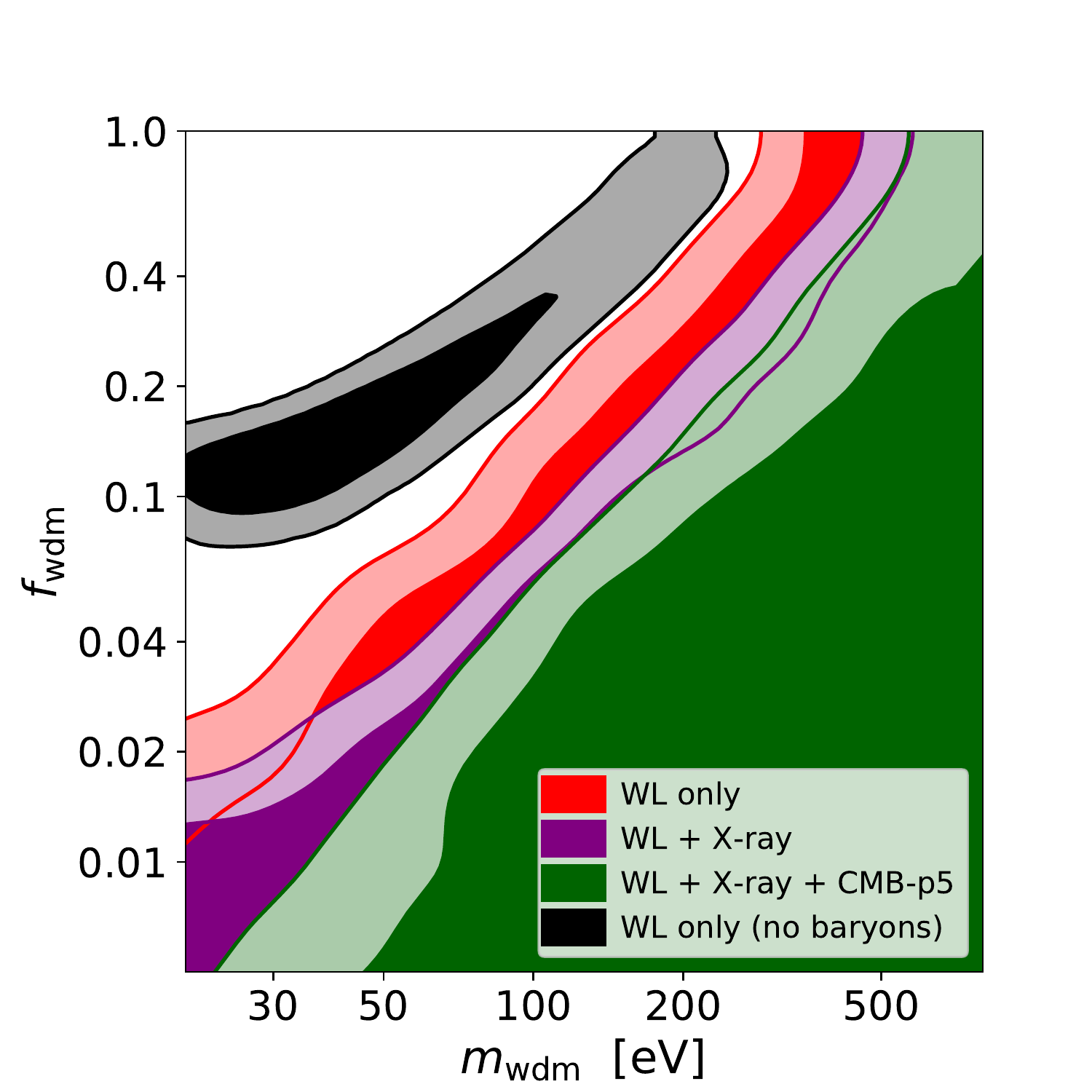}
\caption{Power spectra and likelihood contours of a mixed dark matter cosmology ($\Lambda$MDM), where the dark matter sector consists of a cold and a warm relic. The parameters $f_{\rm wdm}$ and $m_{\rm wdm}$ describe the fraction of warm to total dark matter and the mass of the warm relic. \emph{Left:} Ratio of selected power spectra with different values of $m_{\rm wdm}$ and $f_{\rm wdm}$ relative to the $\Lambda$CDM case ($f_{\rm wdm}=0$). \emph{Right:} Constraints on $m_{\rm wdm}$ and $f_{\rm wdm}$ from different forecast scenarios. All other parameters are marginalised over. The WL only and the WL+X-ray cases with flat, uninformative priors are shown in red and purple. The green constraints correspond to the WL+X-ray scenario with CMB {\tt Planck} priors of the five standard parameters ($n_s$, $h_0$, $\sigma_8$, $\Omega_b$ and $\Omega_m$). The black contours indicate what happens if baryonic effects are ignored.}
\label{fig:contourMDM}
\end{figure}

The left-hand panel of Fig.~\ref{fig:contourMDM} illustrates the relative effect of a $\Lambda$MDM model on the power spectrum at redshift 0.5, assuming two different thermal masses $m_{\rm wdm}=30,\,80$ eV (dashed and solid lines) and three different values for the fraction $f_{\rm wdm}=0.05,\, 0.1,\, 0.15$ (blue, cyan, and magenta). Note that depending on the choice of parameters, the shape of the suppression can be similar to the baryonic suppression, at least for $k$-values below 5 h/Mpc. This could lead to potential degeneracies with baryonic effects, which is something we will investigate in the following.

The right-hand panel of Fig.~\ref{fig:contourMDM} shows the resulting limits from our MCMC analysis for the WL only scenario (red) and for the WL+X-ray scenarios without and with CMB priors from {\tt Planck} (purple and green). The latter only includes CMB priors on the five standard cosmological parameters ($n_s$, $h_0$, $\sigma_8$, $\Omega_b$ and $\Omega_m$) but not on $\Sigma m_{\nu}$, $m_{\rm wdm}$, and $f_{\rm wdm}$.

Fig.~\ref{fig:contourMDM} shows that at small particle masses of $m_{\rm wdm}=50$ eV and below, the amount of warm/hot dark matter can be constrained to be smaller than 4.7 (6.8) and 2.6 (3.7) percent of the total DM budget at the 68 percent (95 percent) confidence level for the WL only and the WL+X-ray case, respectively. This number can be further pushed down to 0.7 (1.7) percent for the WL+X-ray scenario with {\tt Planck} priors. We conclude that for small particle masses, expected limits on the fraction of warm/hot DM relics are significantly tighter than current constraints that are at the 10-20 percent level \citep{Boyarsky:2008xj,Schneider:2014rda,Schneider:2017vfo,Baur:2017stq,Schneider:2018xba} .

At larger masses, however, the expected weak-lensing constraints are not competitive with limits from small-scale clustering. Considering the pure warm DM case ($f_{\rm wdm}=1$), for example, we obtain a limit of $m_{\rm wdm}>300$ eV, 400 eV, and 650 eV (at the 95 percent confidence level) from the WL only case and the WL+X-ray scenarios without and with CMB priors. These limits are more than a factor of 5 below current constraints from the Lyman-$\alpha$ forest \citep[see e.g. Ref.][]{Viel:2013apy}. This is not surprising since weak lensing mainly probes the medium scales of structure formation.

Comparing the WL only results with the ones from the WL+X-ray scenario, we notice that including X-ray gas fractions from eROSITA is expected to provide a significant improvement regarding the expected constrains on $\Lambda$MDM. This can again be explained by the fact that the baryonic and mixed DM power suppression effects are very similar regarding both their shape and amplitude. Constraining baryonic feedback effects with external data, such as cluster gas fractions from X-ray, is therefore important to obtain maximum sensitivity in terms of the dark matter signal.  
 
Finally, the black contours in Fig.~\ref{fig:contourMDM} illustrate what happens if baryonic effects are ignored in the analysis pipeline. In this scenario we falsely predict the presence of a warm/hot relic of $m_{\rm wdm}\lesssim 20-100$ eV that makes up 10-30 percent of the total DM abundance. The case of $f_{\rm wdm}=0$ (assumed for the mock data) is excluded at high significance. This example highlights the need to properly include baryonic feedback effects before probing dark matter models with stage-IV weak-lensing surveys.

\section{Conclusions}\label{sec:conclusions}
Upcoming surveys such as Euclid and LSST will be significantly more accurate and extend to smaller scales compared to current weak-lensing observations. As a consequence, we require a solid theoretical understanding of nonlinear structure formation including the effects of baryonic feedback. This is the second paper in a series of two dedicated to the study of baryonic effects on the cosmic shear power spectrum of a Euclid-like survey. In the first paper, we investigated how baryonic nuisance parameters (defined by the baryonic correction model of S19 \citep{Schneider:2018pfw}) affect the cosmological parameter estimates of a standard $\Lambda$CDM model with massless neutrinos. In this paper, we include the sum of neutrino masses as a free parameter and we combine the weak-lensing data with X-ray mock observations from eROSITA. Furthermore, we investigate three straight-forward extensions of the $\Lambda$CDM model, a dynamical dark energy scenario (wCDM), a $f(R)$ modified gravity model (fRCDM), and a mixed dark matter model ($\Lambda$MDM) with a cold and a warm (or hot) dark matter component. All three scenarios predominantly affect the density perturbations at small scales and are therefore potentially degenerate with the signal from baryonic feedback.

The main findings and conclusions from the present paper are summarised by the following list:
\begin{itemize}
\item Combining the expected cosmic shear power spectrum with X-ray observations of the cluster gas fraction helps to strongly decrease the uncertainties on the 3 baryonic feedback parameters ($M_c$, $\mu$, $\theta_{\rm ej}$). As a consequence, this also leads to modifications of the cosmological parameter contours. The uncertainty on the sum of the neutrino masses ($\Sigma m_{\nu}$) decreases by about 20-30 percent when X-ray data is included. For other parameters, such as $\Omega_m$ or $\sigma_8$, the improvement remains below the 20 percent level. We conclude that constraining baryonic effects with external data results in a noticeable but not very substantial improvement in terms of cosmology. Conversely, this means that properly parametrising baryonic feedback (without knowing the true amplitude and shape of the effect) is sufficient to obtain tight constraints on cosmological parameters within the $\Lambda$CDM framework.

\item Much more significant gains in terms of cosmological posteriors are obtained when WL+X-ray data is combined with information from the CMB. Assuming priors from {\tt Planck} on the five standard parameters ($n_s$, $h_0$, $\sigma_8$, $\Omega_b$ and $\Omega_m$) results in a $\sim70$ percent decrease in the uncertainty of the sum of the neutrino masses ($\Sigma m_{\nu}$) compared to the WL only scenario. The errors of the other cosmological parameters are also reduced significantly below both the WL only scenario and the current limits from {\tt Planck}.

\item The equation-of-state parameters ($w_0$, $w_a$) of the dynamical dark energy model wCDM can be constrained to $\Delta w_0=0.2$ and $\Delta w_a=0.84$ (at the 95 percent confidence level) with the cosmic shear power spectrum alone. Adding X-ray data only leads to a moderate decrease of uncertainty below the 20 percent level. We therefore conclude that simply parametrising baryonic feedback is a sufficient strategy for the wCDM scenario as well.

\item The \citet{Hu:2007nk} $f(R)$ gravity (fRCDM) leads to an enhancement of the power spectrum at small scales which is similar in shape to the baryonic suppression effect, making it a particularly interesting model to study in the context of the present paper. Our forecast analysis reveals that the weak-lensing scenario alone results in a limit of $|f_{R0}|<4\times10^{-6}$ at the 95 percent confidence level. Adding eROSITA-type X-ray leads to $|f_{R0}|<1.6\times10^{-6}$, which is a substantial improvement revealing the necessity to further constrain feedback effects in the context of modified gravity.

\item The mixed dark matter scenario ($\Lambda$MDM) with both a cold and warm/hot relic leads to a power suppression that is very similar in shape and amplitude to the baryonic suppression effect. This means that $\Lambda$MDM is a particularly interesting case to study in combination with baryonic feedback. We find that for a very warm relic with $m_{\rm wdm}<50$ eV, the fraction of warm to total dark matter can be constrained to below 7 percent (at the 95 percent confidence level) with the cosmic shear power spectrum alone. Adding X-ray gas fractions from eROSITA will push this limit down to 3.5 percent which is a significant improvement. An even stronger constraint of 1.7 percent is obtained when {\tt Planck} priors are assumed for the five standard cosmological parameters ($n_s$, $h_0$, $\sigma_8$, $\Omega_b$ and $\Omega_m$). Hence, the $\Lambda$MDM scenario is another example where additional external constraints on baryonic feedback help to significantly decrease uncertainties on fundamental model parameters. 

\item Ignoring baryonic effects in the cosmological prediction pipeline leads to very significant biases of the posterior distributions visible in all parameters. While the existence of large biases has already been pointed out in Paper I, we show here that this remains true in the context of cosmological models with more free model parameters. For example, ignoring baryonic effects results in a neutrino mass estimate in conflict with current solar and atmospheric neutrino experiments. Furthermore, it leads to highly significant false detections of either dynamical dark energy or a subdominant warm/hot dark matter component.
\end{itemize}
Summarising the results of Paper I and II, we want to stress that including baryonic effects on the cosmic shear power spectrum is crucial for future stage-IV weak-lensing surveys like Euclid and LSST. However, once these effects are included in a parametrised form, we find surprisingly tight constraints on cosmological parameters by simply marginalising over baryonic effects. While additional data (such as X-ray gas fractions) helps to further decrease the errors, the improvement is only of the order of 30 percent or less for parameters of the $\Lambda$CDM model. The situation can be different for models beyond $\Lambda$CDM, where better constraints on baryonic effects may lead to significantly tighter limits on fundamental model parameters.

Note that the present paper is limited to the tomographic shear power spectrum. It will be interesting to see if the above conclusions remain valid for higher-order statistics \citep{Foreman:2019ahr,Barreira:2019ckp} or weak-lensing peaks \citep{Weiss:2019jfx,Coulton:2019enn} which are sensitive to non-Gaussian features of the cosmic density field.

\section*{Acknowledgments}
The authors would like to thank Uwe Schmidt for computational support as well as Adam Amara, Raphael Sgier, and Federica Tarsitano for many helpful scientific discussions. This work has been supported by the Swiss National Science Foundation via the project numbers PZ00P2\_161363 and PCEFP2\_181157. AR wants to thank the KIPAC institute at Stanford University where part of this work has been completed. Some of the plots have been created using the {\tt conrer.py} software package \citep{Foreman-Mackey:2016aaa}.


\bibliographystyle{unsrtnat}
\bibliography{ASbib}

\appendix

\begin{figure}[tbp]
\centering
\includegraphics[width=0.98\textwidth,trim=0.1cm 0.1cm 0.5cm 0.4cm,clip]{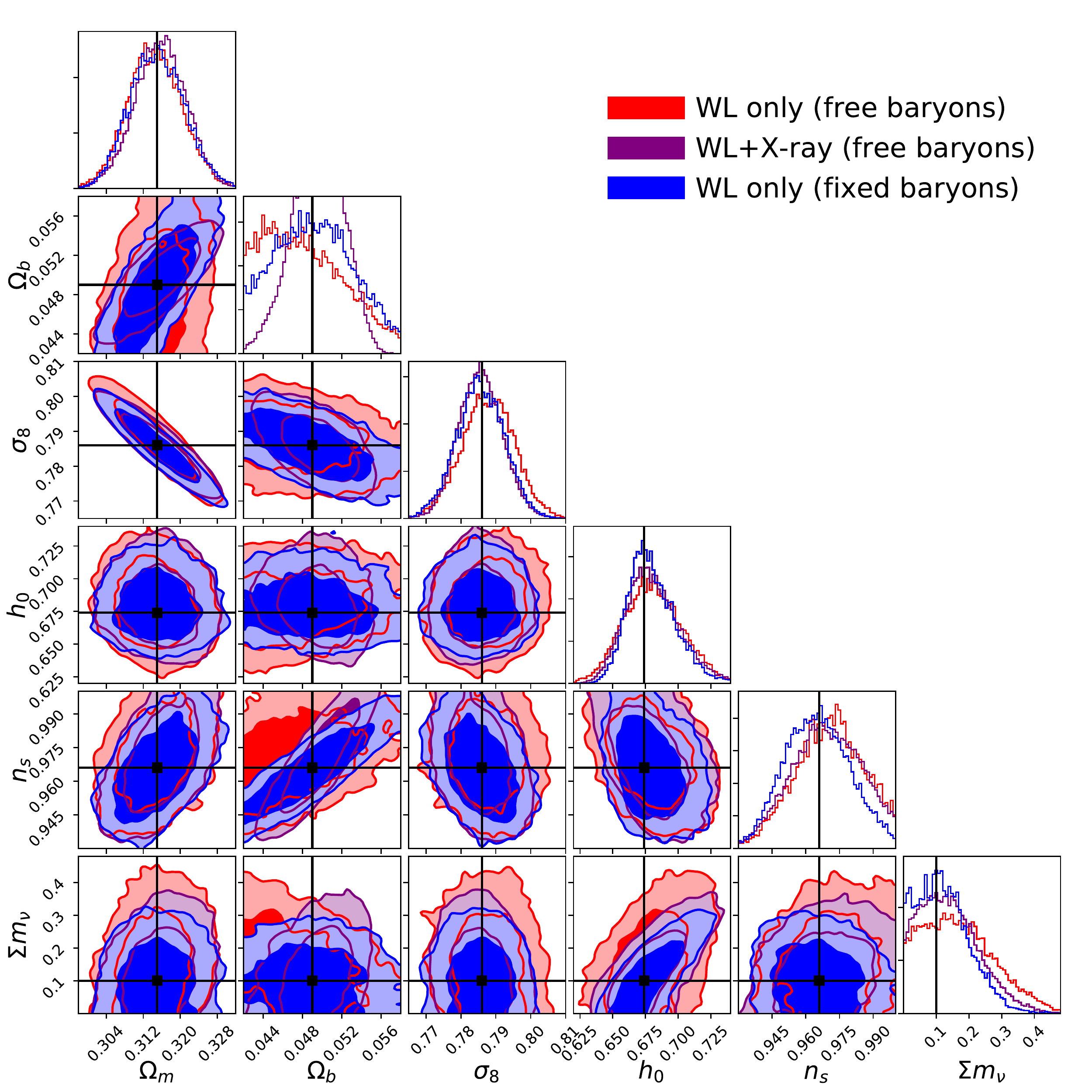}
\caption{Posterior contours of all cosmological parameters assuming a $\Lambda$CDM model with massive neutrinos. The red and blue contours are obtained from the Euclid-like tomographic shear power spectrum alone assuming free or fixed baryonic parameters. The purple contours also include mock data from X-ray gas fractions expected by eROSITA. The true parameter values of the mock data is shown as black lines.}
\label{fig:contours_free_fixed}
\end{figure}

\section{Combining weak lensing and X-ray versus fixing baryonic parameters}\label{app:fixedbaryons}
In Sec.~\ref{sec:likelihoodanalysis} of the main text, we have shown that it is possible to simultaneously constrain baryonic and cosmological parameters using the cosmic shear power spectrum together with the cluster gas fraction from X-ray data. While the former is mainly constraining cosmology, the latter narrows down the uncertainty on baryonic feedback, allowing to further tighten cosmological parameter estimates. In this appendix, we investigate how well the WL+Xray combination used in the main text compares to the idealistic case of full knowledge of baryonic processes. We do this by fixing the baryonic parameters to their true values used in the mock data. This corresponds to the assumption that we fully understand the effect of baryonic feedback on the cosmic density field at scales relevant for weak lensing. 

The blue contours in Fig.~\ref{fig:contours_free_fixed} illustrate the cosmological posterior contours of the six $\Lambda$CDM parameters, assuming the baryonic parameters to be fixed at $\log M_c=13.8$, $\mu=0.21$, and $\theta_{\rm ej}=4$. The other contours in red and purple correspond to the results of the WL only and WL+X-ray scenarios with free baryonic parameters introduced in the main text (see scenario i and ii in Sec.~\ref{sec:likelihoodanalysis}).

The main point to take home from Fig.~\ref{fig:contours_free_fixed} is that the purple posteriors are close to the optimum case shown in blue. For most cosmological parameters, the uncertainties decrease by less than 10 percent when going from the combined WL+X-ray case to the idealistic scenario of fixed baryonic parameters. Only the sum of the neutrino masses ($\Sigma m_{\nu}$) shows a more significant improvement of about 25 percent, decreasing from $\Delta{\Sigma m_{\nu}}=0.126$ to $\Delta{\Sigma m_{\nu}}=0.094$ eV at 68 precent confidence level.

Note furthermore that in  Fig.~\ref{fig:contours_free_fixed} the posterior of cosmic baryon abundance ($\Omega_b$) is significantly tighter when X-ray data is included (purple) compared to the weak-lensing only cases with both free and fixed baryonic parameters (red and blue). This can be explained by the fact that next to constraining baryonic effects, the X-ray data also carries some information about cosmology. Most notably, the X-ray gas fractions are sensitive to the ratio of gas to total matter in the universe, which is closely related to the cosmic baryon abundance ($f_b=\Omega_b/\Omega_m$).

\begin{figure}[tbp]
\centering
\includegraphics[width=0.325\textwidth,trim=0.2cm 0.1cm 1.4cm 0.4cm,clip]{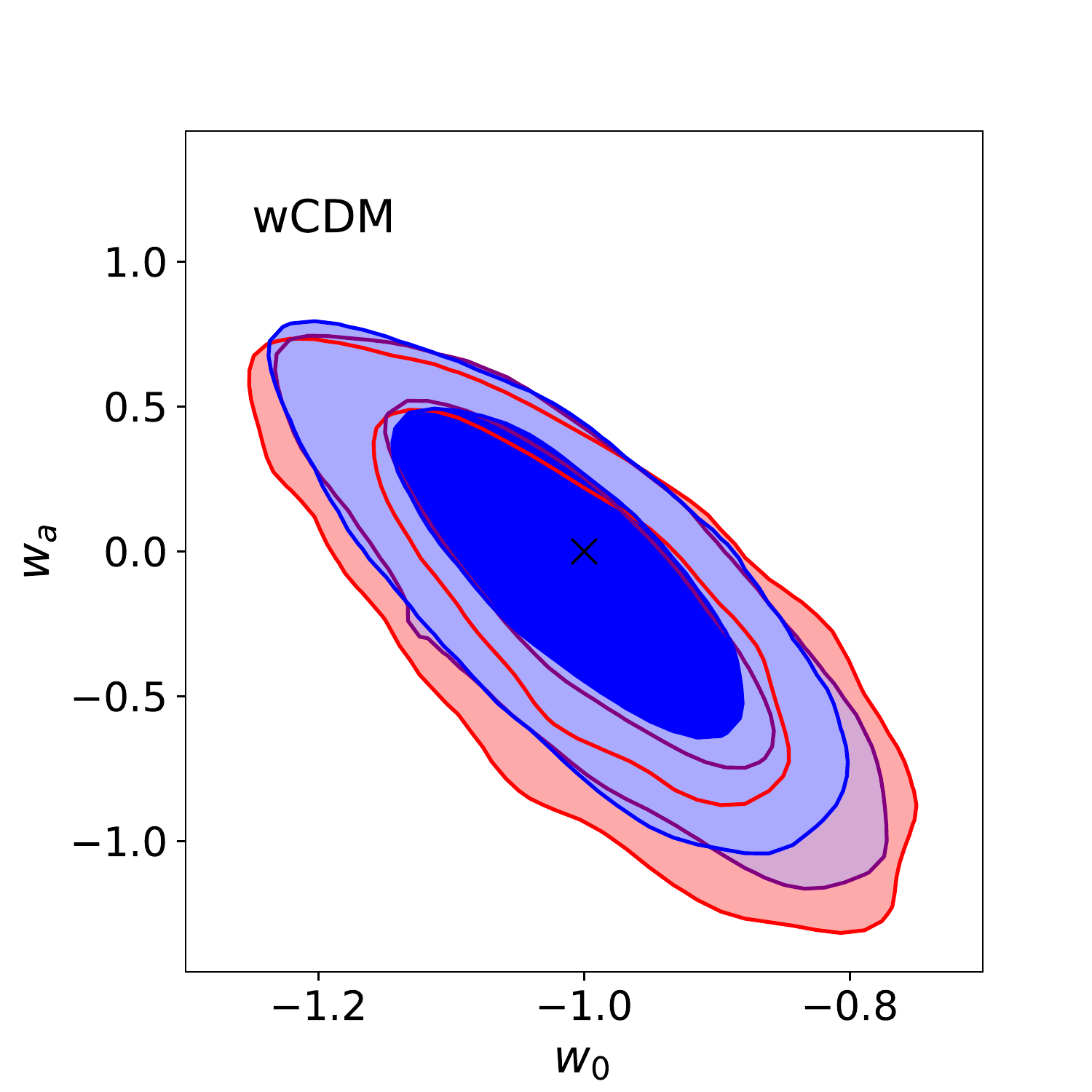}
\includegraphics[width=0.325\textwidth,trim=0.2cm 0.1cm 1.4cm 0.4cm,clip]{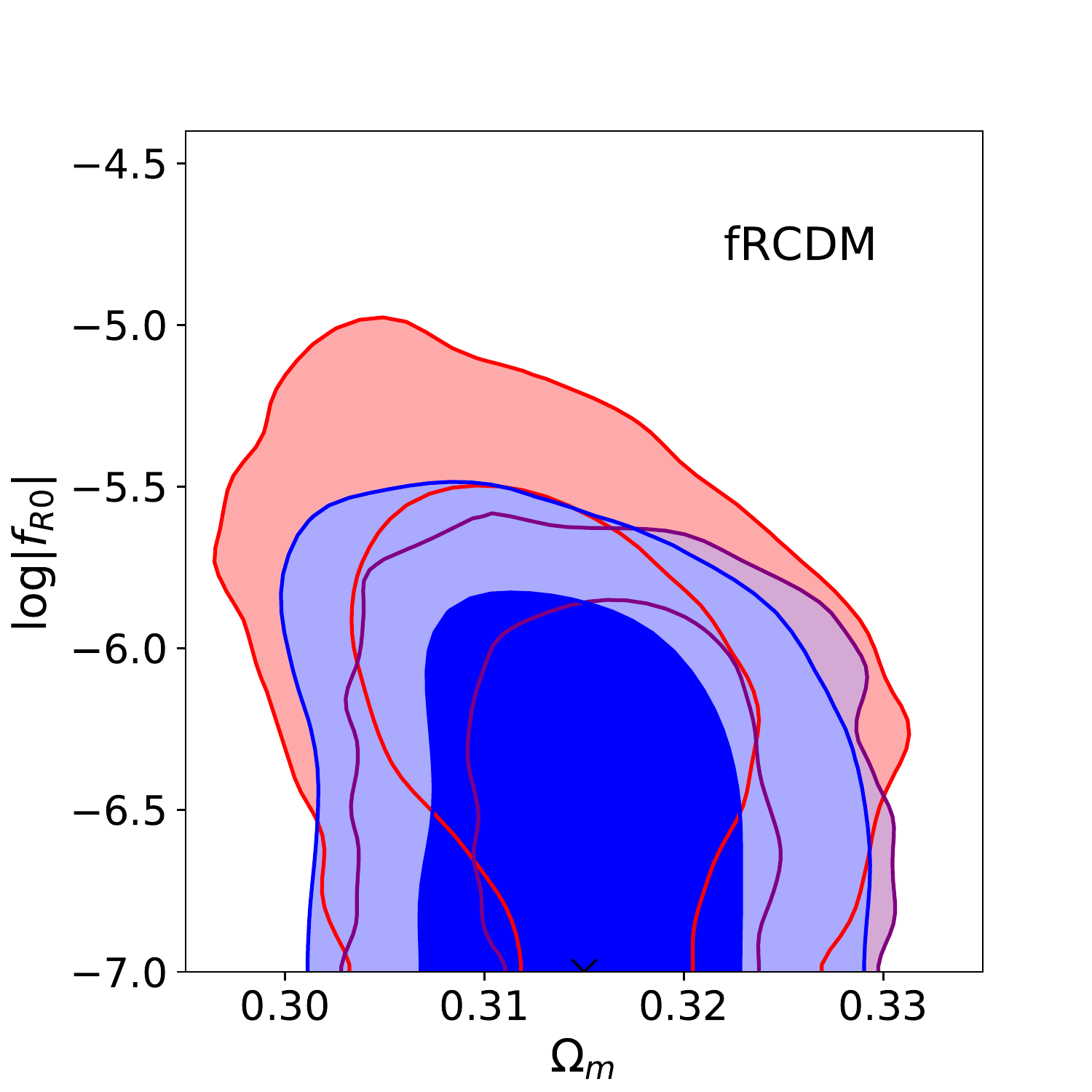}
\includegraphics[width=0.325\textwidth,trim=0.2cm 0.1cm 1.4cm 0.4cm,clip]{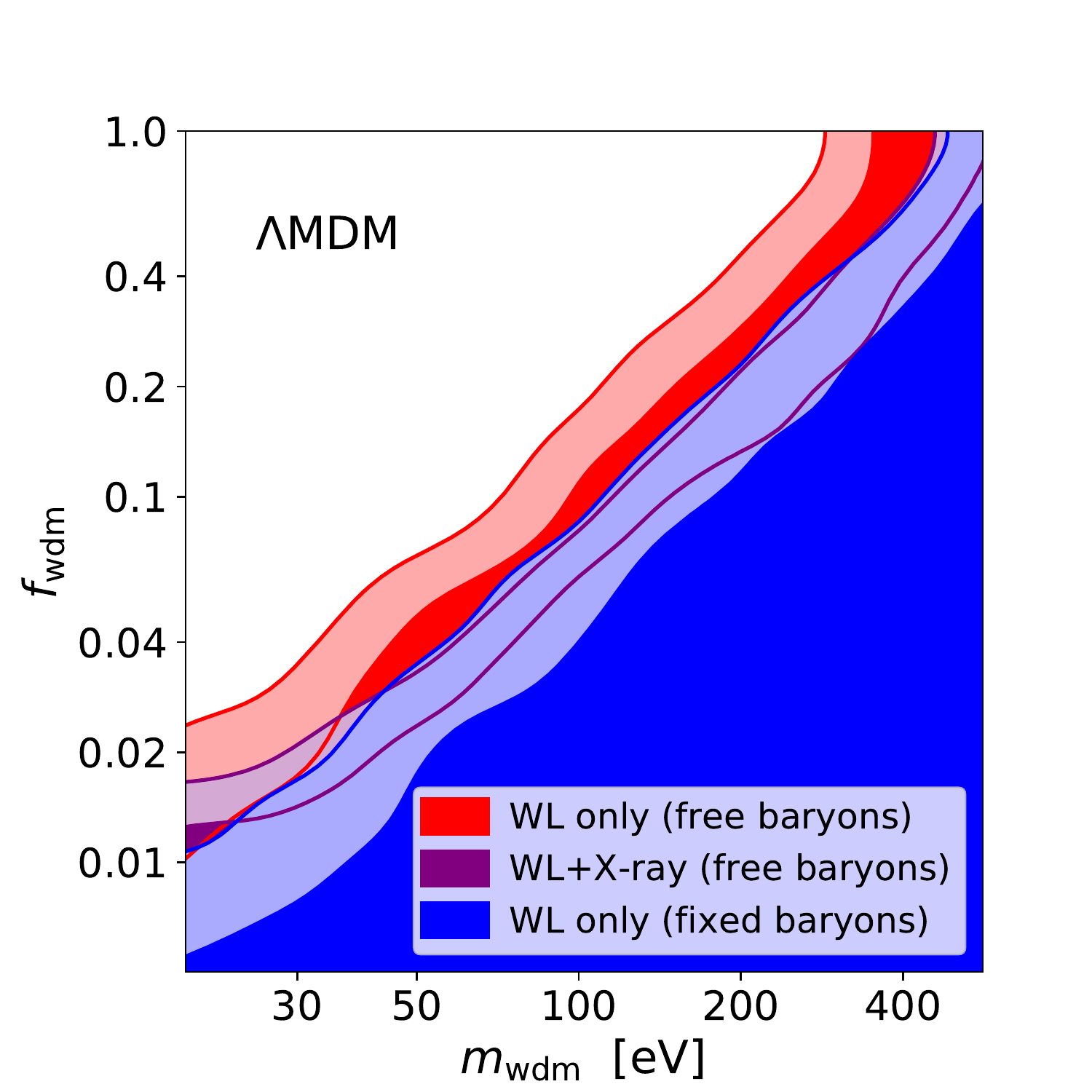}
\caption{Posterior contours of key parameters for wCDM (dynamical dark energy), fRCDM (modified $f(R)$ gravity), and $\Lambda$MDM (mixed dark matter) cosmologies. The red and blue contours show the WL only cases with free and fixed baryonic parameters, while the purple contours correspond to the combined case including both weak lensing and X-ray data.}
\label{fig:contours_free_fixed_beyondLCDM}
\end{figure}

The results obtained so far can be generalised to models beyond $\Lambda$CDM. In Fig.~\ref{fig:contours_free_fixed_beyondLCDM} we show the posterior contours for the wCDM (left), fRCDM (centre), and $\Lambda$MDM cosmologies (right) discussed in Sec.~\ref{sec:extensions}. As already shown in the main text (see Figs.~\ref{fig:contourW0WA}-\ref{fig:contourMDM}), there is a noticeable gain when adding external data from X-ray gas fractions, i.e when going from the WL only (red) to the WL+X-ray scenario (purple). Fig.~\ref{fig:contours_free_fixed_beyondLCDM} furthermore shows that for all three model extensions, fixing the baryonic parameters to their true values does only lead to a moderate further improvement compared to the WL+X-ray case. 

For wCDM, the Figure of Merit of the scenario with fixed baryonic parameters is at ${\rm FoM}=32$. This is only slightly larger than the ${\rm FoM}=27$ obtained for the WL+X-ray case (scenario iii). Regarding the fRCDM, there is no improvement of the limit on $|f_{\rm R0}|$, while for the $\Lambda $MDM model, an improvement is visible but remains small. Note that in terms of cosmological parameter contours, it is more promising to include further information from the CMB than to find even better ways to constrain baryonic effects. This statement is both true for $\Lambda$CDM and for all extensions considered in the present paper.

\begin{figure}[tbp]
\centering
\includegraphics[width=0.85\textwidth,trim=1.9cm 1.7cm 2.2cm 1.2cm,clip]{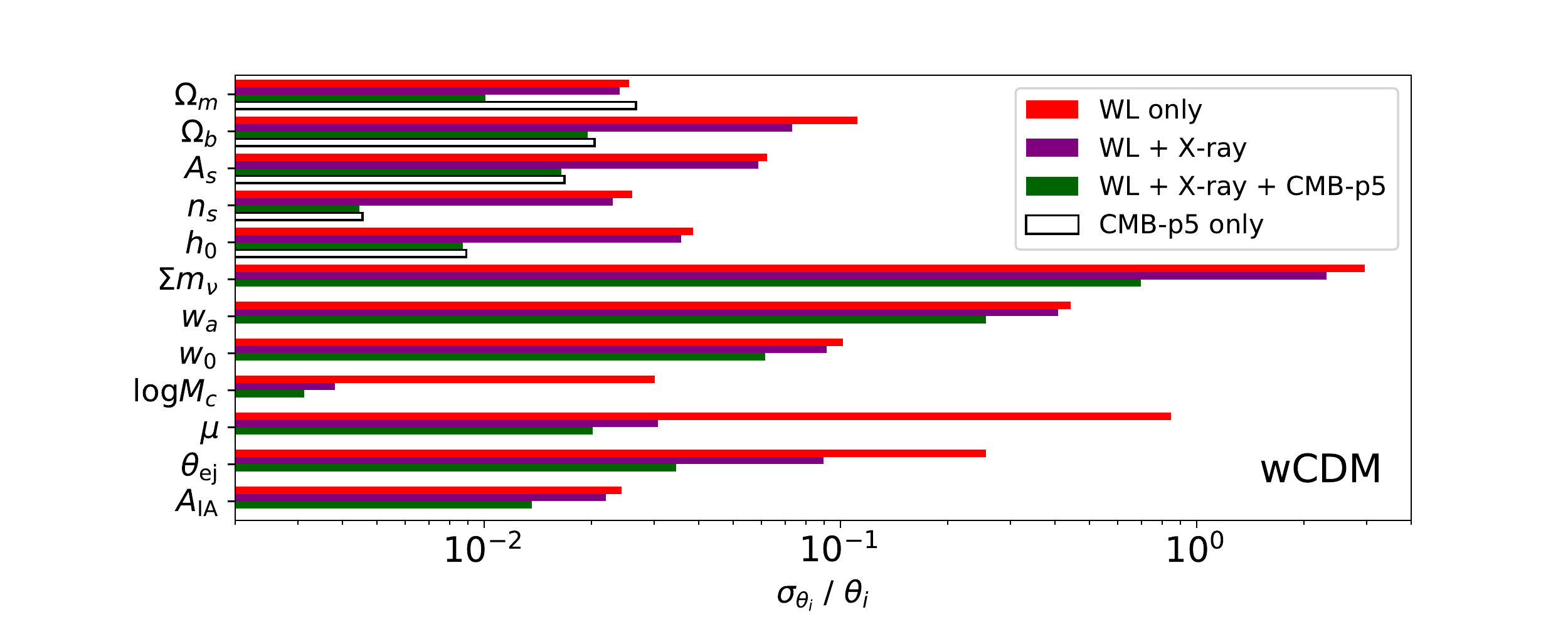}
\includegraphics[width=0.85\textwidth,trim=1.9cm 1.7cm 2.2cm 1.2cm,clip]{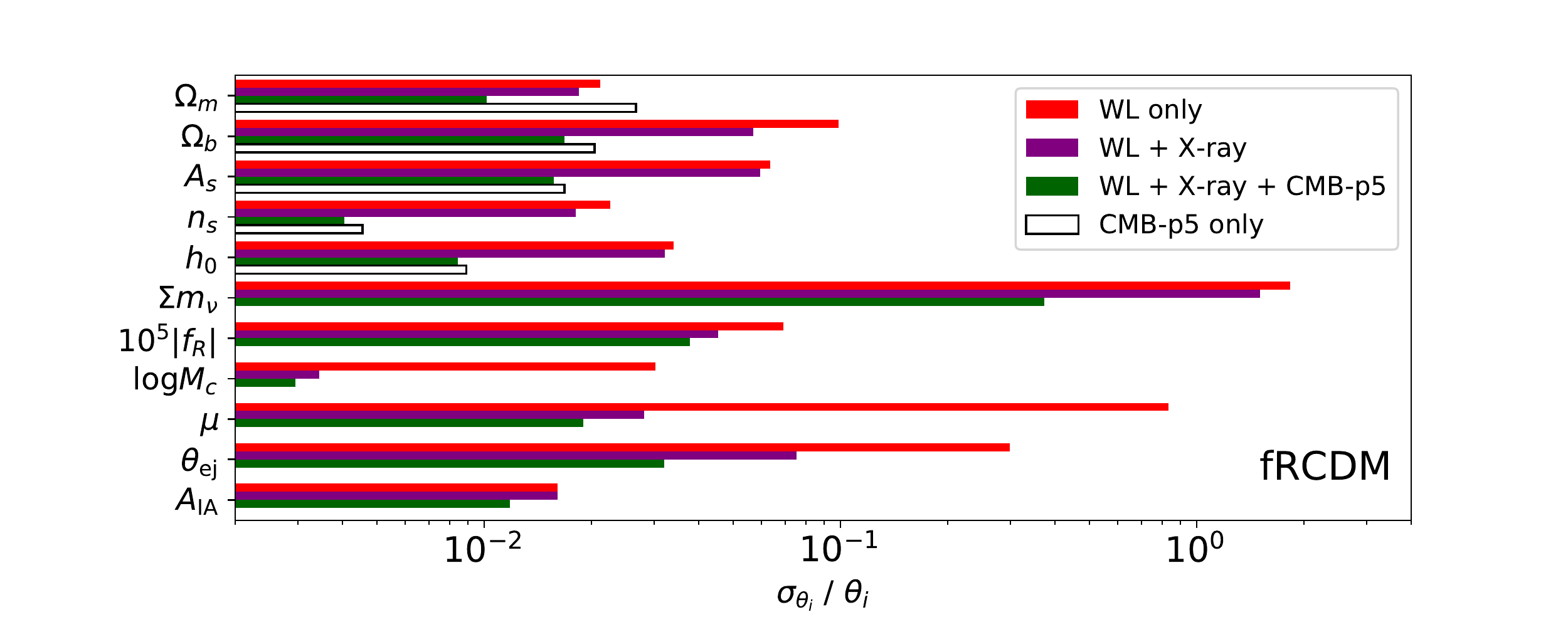}
\includegraphics[width=0.85\textwidth,trim=1.78cm 0.2cm 2.32cm 1.2cm,clip]{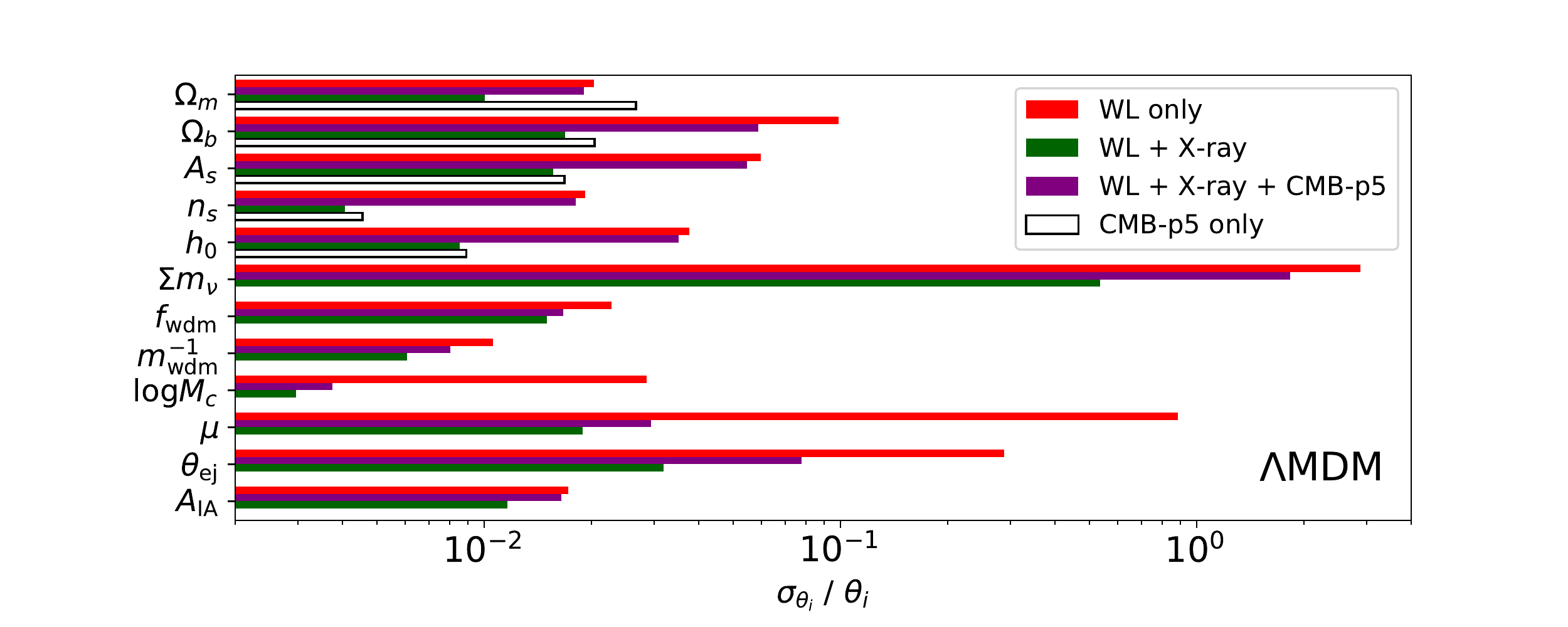}
\caption{Marginalised errors ($\sigma_{\theta_i}$) of all model parameters (at 68 percent confidence level) for the wCDM, fRCDM and $\Lambda$MDM scenarios (top to bottom panels) normalised by the fiducial value of the mock ($\theta_i$). For all parameters with a fiducial value of zero ($w_a$, $|f_R|$, $m_{\rm wdm}^{-1}$, and $f_{\rm wdm}$), we set $\theta_i=1$. Regarding the $\Lambda$MDM scenario, note that $m_{\rm wdm}^{-1}$ is provided at $f_{\rm wdm}=0.1$ and $f_{\rm wdm}$ at $m_{\rm wdm}=100$ eV. The WL only and the WL+X-ray results with flat, uninformative priors are shown as red and purple bars. The green bars correspond to the WL+X-ray scenario with Gaussian priors on $n_s$, $h_0$, $A_s$, $\Omega_b$ and $\Omega_m$ (but not $\Sigma m_{\nu}$) from the CMB experiment {\tt Planck} \citep{Aghanim:2018eyx}. The black empty bars provide the size of these CMB priors.}
\label{fig:sigma_extendedcosmologies}
\end{figure}

\section{More information about extended cosmologies}\label{app:extendedcosmologies}
In this appendix we provide more details about the forecast study of the wCDM, fRCDM, and $\Lambda$MDM cosmologies introduced in Sec.~\ref{sec:extensions}. While we only focused at the most relevant key parameters in the main text, we now discuss the marginalised errors of all individual parameters that are varied during the MCMC sampling.

In Fig.~\ref{fig:sigma_extendedcosmologies} we plot the individual one-sigma errors ($\sigma_{\theta_i}$) of all parameters (i) divided by the fiducial values of the mock ($\theta_i$) in the three cosmologies. For the parameters where the fiducial value of the mock is zero (such as $w_a$, $\log f_R$, $m_{\rm wdm}^{-1}$, and $f_{\rm wdm}$), the corresponding error is divided by one instead, showing absolute instead of relative posteriors in these cases. Regarding the $\Lambda$CDM scenario, the posteriors of $m_{\rm wdm}$ and $f_{\rm wdm}$ correspond to one-sided constraints that depend on each other. In order to provide meaningful one-sigma posteriors, we therefore plot the constraint on $m_{\rm wdm}^{-1}$ at a fixed value of $f_{\rm wdm}=0.1$ and the constraint on $f_{\rm wdm}$ at $m_{\rm wdm}=100$ eV.

Compared to the $\Lambda$CDM case illustrated in Fig.~\ref{fig:sigma_wlonly_wlxray} of the main text, Fig.~\ref{fig:sigma_extendedcosmologies} shows slightly degraded posterior constraints. This is not surprising as the three beyond-$\Lambda$CDM scenarios come with one or two additional model parameters. Particularly noteworthy is the change regarding the sum of the neutrino masses, where we observe a significant weakening of the constraints for the wCDM and $\Lambda$MDM, but not the fRCDM cosmology.

\end{document}